\documentclass[aps,prc,twocolumn,amsmath,amssymb,floatfix,superscriptaddress]
{revtex4}
\usepackage{mathptmx}
\usepackage{CJK}
\usepackage{graphicx}
\usepackage{bm}
\usepackage{dcolumn}
\usepackage{multirow}
\usepackage{xcolor}
\usepackage{ulem}
\usepackage[
breaklinks,pdfstartview=FitH,CJKbookmarks=true,
bookmarksnumbered=true,bookmarksopen=true,
colorlinks=true,linkcolor=blue,urlcolor=blue,anchorcolor=blue,citecolor=blue
]{hyperref}

\def\beq{\begin{equation}} \def\eeq{\end{equation}}
\def\bal#1\eal{\begin{align}#1\end{align}}
\def\bse#1\ese{\begin{subequations}#1\end{subequations}}
\def\non{\nonumber}
\def\ra{\rightarrow}

\def\al{\alpha}
\def\be{\beta}

\def\la{\Lambda}

\def\fm3{\,\text{fm}^{-3}}

\def\mfm5{\;\text{MeV}\,\text{fm}^5}

\def\c12{$^{12}$C}

\def\pnl{\Phi^{(N\la)}\!}
\def\bbe{\bar{\beta}}

\begin{document}

\begin{CJK*}{UTF8}{gbsn}
\title{Shape coexistence in Ne isotopes and hyperon impurity effect on low-lying states}

\author{Huai-Tong~Xue~(薛怀通)} %
\affiliation{
	College of Physics and Electronic Engineering, Nanyang Normal University, Nanyang 473061, China}
\affiliation{
	Department of Physics, East China Normal University, Shanghai 200241, China}

\author{Ji-Wei~Cui~(崔继伟)} %
\affiliation{
	School of Physics and Optoelectronic Engineering, Xidian University, Xi'an 710071, China}

\author{Q.~B.~Chen~(陈启博)} %
\author{Xian-Rong~Zhou~(周先荣)} \email{xrzhou@phy.ecnu.edu.cn}
\affiliation{
Department of Physics, East China Normal University, Shanghai 200241, China}

\author{Hiroyuki Sagawa} 
\affiliation{
Center for Mathematical Sciences, University of Aizu Aizu-Wakamatsu, Fukushima 965-8580, Japan}


\date{\today}

\begin{abstract}

Based on the beyond-mean-field Skyrme-Hartree-Fock model,
we investigate the shape coexistence in Ne isotopes and the effect of 
$\la$ hyperon on the energy level structure in the nuclei.
The up-to-date Skyrme-type $N\la$ interaction SLL4
and the $NN$ interaction SGII are employed.
Low-lying energy spectra of $^{20,22,24,26,28,30,32,34}$Ne,
including the low-lying states with $J\leq 6$,
are predicted, discussed in detail,
and found in good agreement with experimental results.
The electric quadrupole transition rate 
is also examined.
The coexistences of a ground state rotational band and a $\be$ 
vibrational band are 
revealed in $^{20,22,24}$Ne.
Unlike the previously discovered  shrinkage effect of $\la_{s}$ on 
the ground state nuclei, it is found that the $\la_{s}$ may alter 
the excitation mode of the second band by affecting the distribution 
of the collective wave function, thereby causing the $\be$ vibrational 
band transitions to a vibrational band with equidistant energy levels.
\end{abstract}

\maketitle

\section{Introduction}
\label{s:intro}
The shape of a nucleus is one of its most fundamental properties, 
and its exploration across the nuclear landscape provides insight into the 
mechanisms underlying how protons and neutrons are organized~\cite{Garrett2019PRL}.
Nuclear shape coexistence is the phenomenon in which distinct shapes occur 
within the same nucleus and at a similar energy.
Minima in the total nuclear potential energy can be found for shapes 
that include spherical, 
axially symmetric prolate or oblate deformed ellipsoids, 
axially nonsymmetric (triaxial) ellipsoids, etc. 
The lowest minimum is associated with the mass 
and shape of the nucleus. Additional minima are shape isomers and in even-even 
nuclei are manifested as excited $0^{+}$ states~\cite{Aberg1990}.

After it was first proposed in 1956, shape coexistence has been observed in  
heavy and medium-heavy nuclei, as well as light nuclei thanks to 
many experimental~\cite{Garrett2022PLB} and theoretical attempts ~\cite{Heyde2011RMP,Heyde1983PR}.
For example, shape coexistence has been observed in nuclei with Z 
$\sim$ 82~\cite{Bonn1972PLB}, 
Z $\sim$ 50~\cite{Bron1979NPA}, Z $\sim$ 40 and N $\sim$ 60~\cite{Cheifetz1970PRL,Federman1977PLB,Federman1979prc}, 
 $Z \sim$ 64 and  $N \sim$ 90~\cite{Hager2007NPA}, 
 $Z \sim$ 34 and  $N\sim$ 40~\cite{Hamilton1974PRL}, nuclei in the light mass region with $N = Z$, 
and nuclei in the island of inversion with $(N, Z)=(8, 6), (20, 12)$, and $(28, 14)$.
In recent years, an up-to-date view of the experimental manifestation of 
shape coexistence in nuclei and theories predicting its occurrence think that there appears 
to be a possibility that it occurs in all nuclei~\cite{Heyde2011RMP}.


At the same time, various theoretical approaches have been developed 
to describe nuclear shape coexistence, 
including the interacting shell model~\cite{Caurier2005RMP}, 
the Monte Carlo shell model~\cite{Otsuka2001PPNP}, 
the interacting boson model~\cite{Nomura2016JPG}, 
and both nonrelativistic~\cite{Bender2003RMP,Robledo2019JPG} and 
relativistic/covariant~\cite{NIKSIC2011PPNP} density functional theories (DFTs).

Among these approaches, nuclear DFT stands out as the most efficient microscopic 
method capable of offering a unified and consistent description for a wide range 
of nuclei across the nuclear chart~\cite{Bender2003RMP}. Initially designed as a ground-state theory, 
nuclear DFT necessitates expansion beyond the mean-field level to address 
nuclear spectroscopic properties adequately. One viable solution involves 
the utilization of a beyond-mean-field approach, which encompasses 
angular momentum projection (AMP) techniques and the generator coordinate method (GCM). 
These methods are founded on the Skyrme-Hartree-Fock (SHF) approach, 
with collective parameters derived from DFT calculations.

It represents various quantum configurations with distinct spatial arrangements, 
all varying for the lowest energy state. This phenomenon provides an excellent 
platform for studying the interactions among these configurations within a single nucleus.
Recently, Ne isotopes have attracted increasing interests. 
Thus, they have been extensively studied experimentally and theoretically.
Simultaneously, some of the shape evolution and shape coexistence in Ne isotopes 
have been examined in multiple investigations.
In Ref.~\cite{Sagawa04}, the isospin dependence of the shapes 
in Ne isotopes have been studied by analyzing the quadrupole moments 
and electric quadrupole transitions $B(E2)$ utilizing the deformed SHF model.
As reported in Ref.~\cite{Lia2013prc}, the quadruple deformations in Ne isotopes and 
the corresponding $\la$ hypernuclei were investigated by the deformed SHF+BCS model 
including different tensor and pairing forces.
The angular momentum projected generator coordinate method based on 
the Gogny force (D1S) mean field was also applied to investigate 
the quadrupole collectivity in neutron-rich Ne isotopes 
by analyzing potential energy surface~(PESs) and the spectroscopic quadrupole moments~\cite{Rodriguez2003}.
A very strong shape coexistence, which exhibited an oblate ground state 
and a prolate isomeric state with an energy difference of $77$ keV, 
was predicted in $^{24}$Ne.

Meanwhile, the investigation of hypernuclei is also another hot topic in nuclear physics nowadays.
The explorations of hypernuclear systems comprising nucleons and hyperons
can exert far-reaching implications on nuclear physics~\cite{Feliciello2015, Gal16}.
In recent years, nuclear mean-field (MF) energy-density functionals (EDFs)~\cite{Schulze2014PRC,
Xue22,Xue23,chen_2022_epja,Guo2022prc,Liu2024prc,Li2024prc} and 
beyond-MF approaches~\cite{Cui2017prc,Cui2022cpc,Mei2015prc,XJH2019,Xue2024prc} were extended to hypernuclei.

The $\la$ hypernucleus, composed of an ordinary nuclear core and the lightest hyperon, 
provides a unique laboratory to study the $\la N$  interactions.
In addition, $\la$ hyperon can induce multiple interesting effects such as the shrinkage of 
the nuclear size and the stabilization of the binding energys~\cite{Motoba1983PTP, Hiyama1999}, 
the changes in the nuclear cluster structures~\cite{Hiyama1997PTP,Yao2011NPA,Hagino2013NPA}, 
the extension of the neutron drip line~\cite{Vretenar1998prc,Zhou2008prc,Lu2003epja}, 
the appearance of nucleon and hyperon skin or halo~\cite{Hiyama1996prc,Xue22}. Especially in recent years, interplay between the octupolely deformed $^{20}$Ne state and a $\Lambda$ hyperon has been presented in Refs.~\cite{XJH2019,XJH2023}.
Although the impurity effect of hyperons on shape coexistence has been studied 
at the mean field aspect~\cite{Chen2021SCP}, the change of energy spectra due to the interplay between the shape coexistence and the $\Lambda$ hyperon has not been carefully explored. Such researches need the restoration of rotatinal symmetry and interaction between different shapes, which has been successfully realized in our recent work~\cite{Xue2024prc}.

The purpose of this paper is to use the beyond SHF model with AMP and GCM calculations to study the neon hyperisotopes from  $A= 20$ up to  $A= 34$ and to investigate the impurity effect of $\la$ hyperon on shape coexistence.

This paper is organized as follows. Sec.~\ref{s:theo} reviews briefly 
the formalism of the beyond-SHF model, Sec.~\ref{s:results} presents the results 
and discussions and in Sec.~\ref{s:summary} we summarize the work.

\section{Theoretical framework}
\label{s:theo}

The hypernuclear MF wave function
obtained from a SHF calculation with a quadrupole constraint is given by:
\beq\label{e:wav}
\big| \pnl(\be) \big\rangle =
\big| \Phi^N(\be) \big\rangle \otimes \big| \Phi^\la \big\rangle \:,
\eeq
where $\big|\Phi^N(\be)\big\rangle$ and $\big|\Phi^{\la}\big\rangle$
are intrinsic wave functions of the nuclear core and of the $\la$ hyperon,
respectively.
More specifically,
the hyperon wave function for single-$\la$ hypernuclei is
\beq
\big|\Phi^\la\big\rangle = \varphi_s(\la) \:,
\eeq
and
\beq
\left| \Phi^N(\be) \right\rangle =
\prod_{k>0} \left( u_k + v_k a_k^+ a_{\bar k}^+ \right)
| \text{HF} \rangle \:
\eeq
is a BCS state obtained from the nuclear SHF+BCS calculation
with density-dependent delta interaction (DDDI)~\cite{Bender2000epja},
constrained to an
axially-deformed shape given by the deformation parameter $\be$,
which is proportional to the quadrupole moment,
\beq
\be = \frac{4\pi}{3 A_c R_c^2}
\big\langle \Phi^N(\be) \big|r^2 Y_{20}\big| \Phi^N(\be) \big\rangle \:,
\eeq
where $A_c$ is the mass number of the core nucleus, and
$R_c \equiv 1.2 A_c^{1/3}\;$fm.


In the GCM, The hypernuclear states are given by a superposition
of projected MF wave functions onto exact angular momentum $J$:
\beq\label{e:psi}
\left|\Psi_\al^{JM}\right\rangle =
\sum_\be F_\al^J(\be) \hat{P}_{MK}^J
\big|\pnl(\be)\big\rangle \:,
\eeq
where $F_\al^J(\be)$ is a weight function,
and $\hat{P}_{MK}^J$ is the AMP operator,
with $K=K_\text{core}+K_\la$ representing the projection
of angular momentum $J^\pi$ onto the intrinsic $z$ axis.

To obtain the eigenstate $\left|\Psi_\al^{JM}\right\rangle$,
each weight $F_\al^J(\be)$ in Eq.~(\ref{e:psi})
is determined by the Hill-Wheeler-Griffin (HWG) equation~\cite{Peter1980},
\beq
\sum_{\be'}
\Big[ H_{KK}'^J(\be,\be') - E_\al^J N_{KK}^J(\be,\be') \Big]
F_\al^J(\be') = 0 \:,
\label{e:hwg}
\eeq
in which the corrected Hamiltonian and norm elements are given by
\bal
H_{KK'}'^J(\be,\be') &=
\big\langle \pnl(\be') \left| \hat{H}' \hat{P}_{KK'}^J \right|
\pnl(\be) \big\rangle \:,
\\
N_{KK'}^J(\be,\be') &=
\big\langle \pnl(\be') \left| \hat{P}_{KK'}^J \right|
\pnl(\be) \big\rangle \:.
\label{e:nkk}
\eal
The corrected Hamiltonian $\hat{H}'$ is defined as
\beq
\hat{H}' = \hat{H}
- \lambda_p (\hat{N}_p-Z) - \lambda_n (\hat{N}_n-N) \:,
\eeq
where the Hamiltonian $\hat{H}$ is determined by the hypernuclear EDF,
and the last two terms account for the fact
that the projected wave function
does not provide the correct number of particles on average~\cite{Bonche1990npa},    
The projected energy curve onto a specific angular momentum is derived as
\beq
E_{JK}(\be) = {H'^J_{KK}(\be,\be) \over N^J_{KK}(\be,\be)} \:.
\label{e:ejk}
\eeq

Since the projected states do not form an orthogonal basis,
$F_\al^J(\be)$
are nonorthogonal functions~\cite{Mei2015prc},
and orthogonal collective wave functions are constructed as
\beq
g_\al^J(\be) = \sum_{\be'}
\big[ \mathcal{R}^\frac12 \big]^J\!\!(\be,\be') F_\al^J(\be') \:,
\label{e:g}
\eeq
which are weights of the natural states in the collective subspace~\cite{Peter1980},
and where
\beq
\big[ \mathcal{R}^{\frac12} \big]^J\!\!(\be,\be') =
\sum_k \sqrt{n_k} w_k(\be) w_k^*(\be')
\eeq
with the eigenfunctions $w_k$ and eigenvalues $n_k$ of the norm operator,
Eq.~(\ref{e:nkk}), in the projected space.
The average deformation
\beq
\bbe_\al^J = \sum_\be \left| g_\al^J(\be) \right|^2 \be
\label{e:bb}
\eeq
reflects the shape of the dominant configurations in the ground
or excited state and indicates the band structure~\cite{Bender2006prc}.

 To avoid the cancellation of two dominant configurations with different shapes(oblate/prolate), 
another average deformation is introduced in this paper as below,

\beq
\overline{|\be_\al^J|} = \sum_\be \left| g_\al^J(\be) \right|^2 |\be|
\label{e:bb2}
\eeq

Given the weight function $F_\al^J(\be)$,
the root-mean-square (rms) radius is defined as
\beq
R_\text{rms}^{J\al} =
\sqrt{ \sum_{\be\be'} {F_\al^J(\be')}^* F_\al^J(\be)
	\big\langle \pnl(\be') \big| r^2 \hat{P}_{KK}^J
	\big| \pnl(\be) \big\rangle } 
\eeq
with $r^2 = \frac1A \sum_k r_k^2$,
and the reduced $E2$ transition rate is derived as
\beq
B(E2,J_\al^+ \to J_{\!\al'}'^+) = \frac{1}{2J+1}
\left|\langle J_{\!\al'}'^+ || \hat{Q}_2 || J_\al^+ \rangle\right|^2 \:,
\label{e:BE2}
\eeq
where the reduced matrix element is
\bal
& \langle J_{\!\al'}'^+ || \hat{Q}_2 || J_\al^+ \rangle =
\sqrt{2J'+1}
\label{e:QMATRX}
\\\non
& \times \sum_{M\mu\be\be'} {F^{J'}_{\al'}(\be')}^*
F^J_\al(\be) \;C^{J'K'}_{J M 2\mu}
\big\langle \pnl(\be') \big| \hat{Q}_{2\mu}\hat{P}^J_{MK}
\big| \pnl(\be) \big\rangle \:,
\eal
in which
$C^{J'K'}_{JM2\mu}$ denotes the Clebsh-Gordon coefficients, and
$\hat{Q}_{2\mu} = \sum_k e_k r_k^2 Y_{2\mu}(\varphi_k,\theta_k)$
is the electric quadrupole transition operator~\cite{Dobaczewski09},
where $e_k$ is the charge of the $k$th nucleon
and $r_k$ is its position relative to the center of mass of the nucleus.
Bare charges are used in this calculation
(i.e., $e_p=e$ and $e_n=e_\la=0$).



\begin{figure*}[htb]
	\includegraphics[scale=0.6]{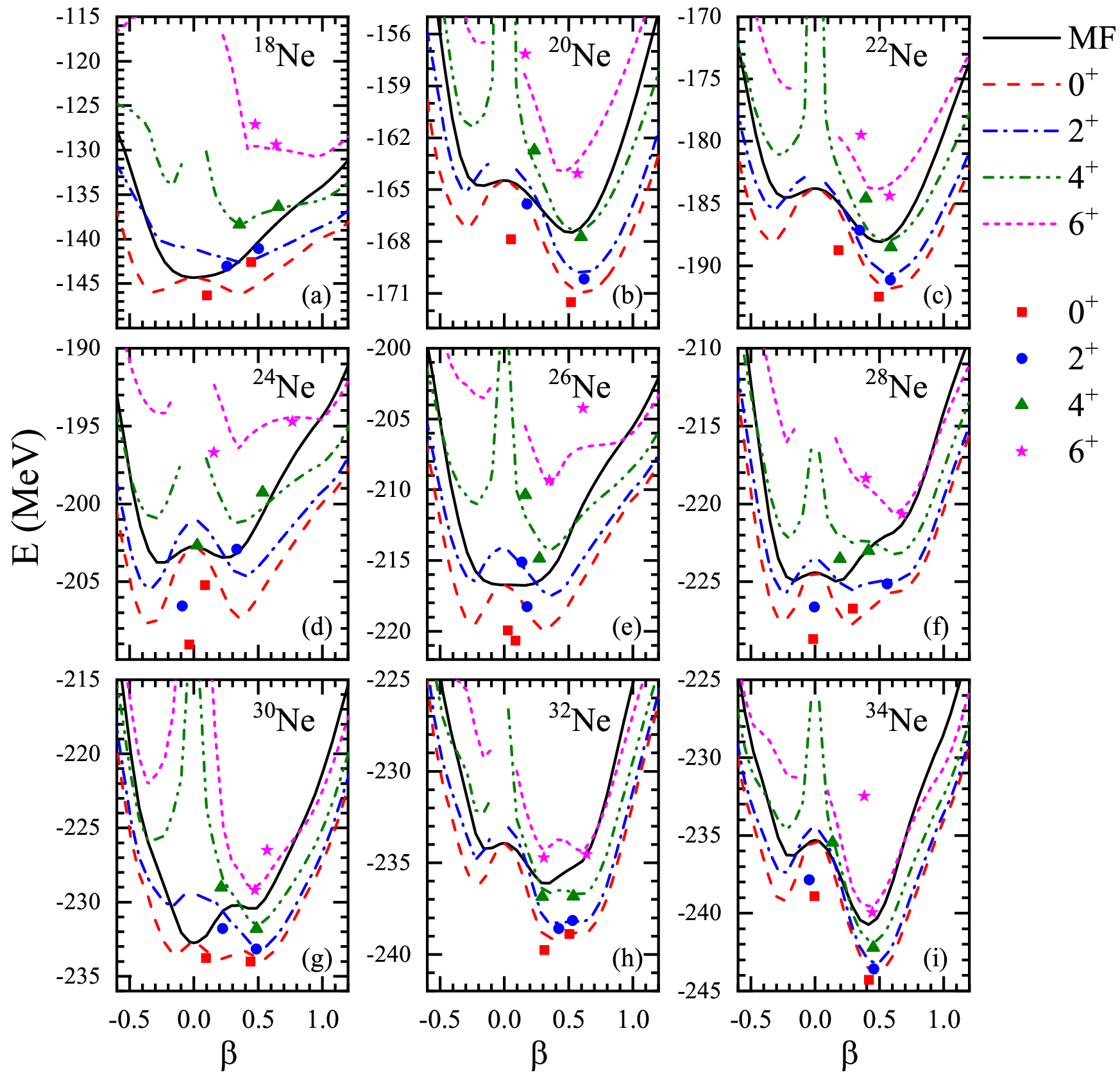}
	\vskip-3mm
	\caption{(Color online) Projected PESs, E($\beta$, $J$), and the GCM energy levels of Ne-isotopes.
		The angular momentum and parity for each projected PES are given in the legend,
		and the mean-field PES labeled by MF is also shown for comparision.
		The solid bullets indicate the GCM energy levels, which are plotted at their average deformation. 
	}
	\label{f:eps-amp}
\end{figure*}

\begin{figure*}[htb]
	\includegraphics[scale=0.6]{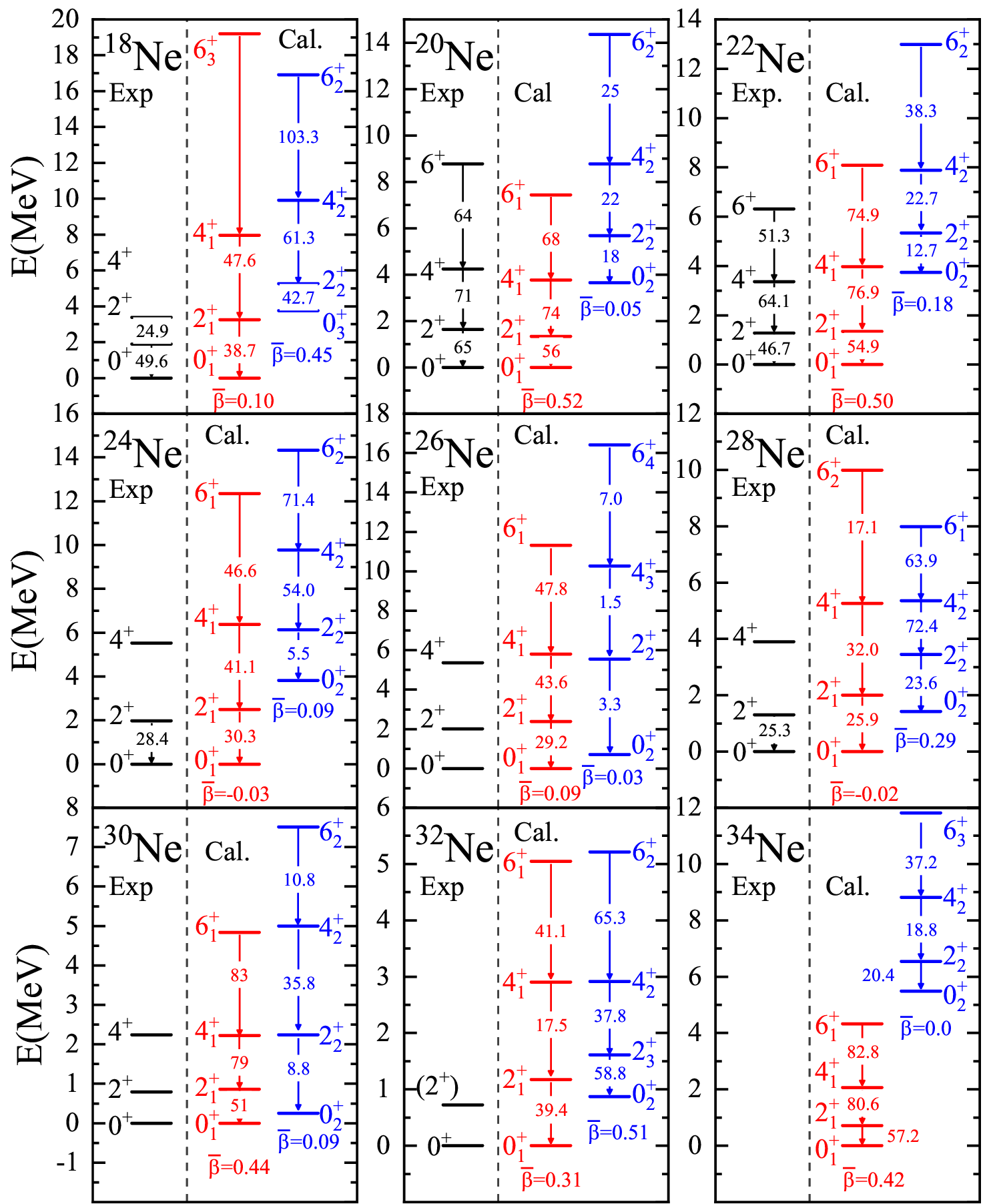}
	\caption{
		(Color online) The spectrum of collective states for $^{20,22,24}$Ne 
		as seen in the experiment ($1st$ column) and as obtained from 
		our calculation ($2$nd column for g.s. band, $3$rd column for another band, 
	}
	\label{f:level}
\end{figure*}

\begin{figure*}[htb]
	\includegraphics[scale=0.55]{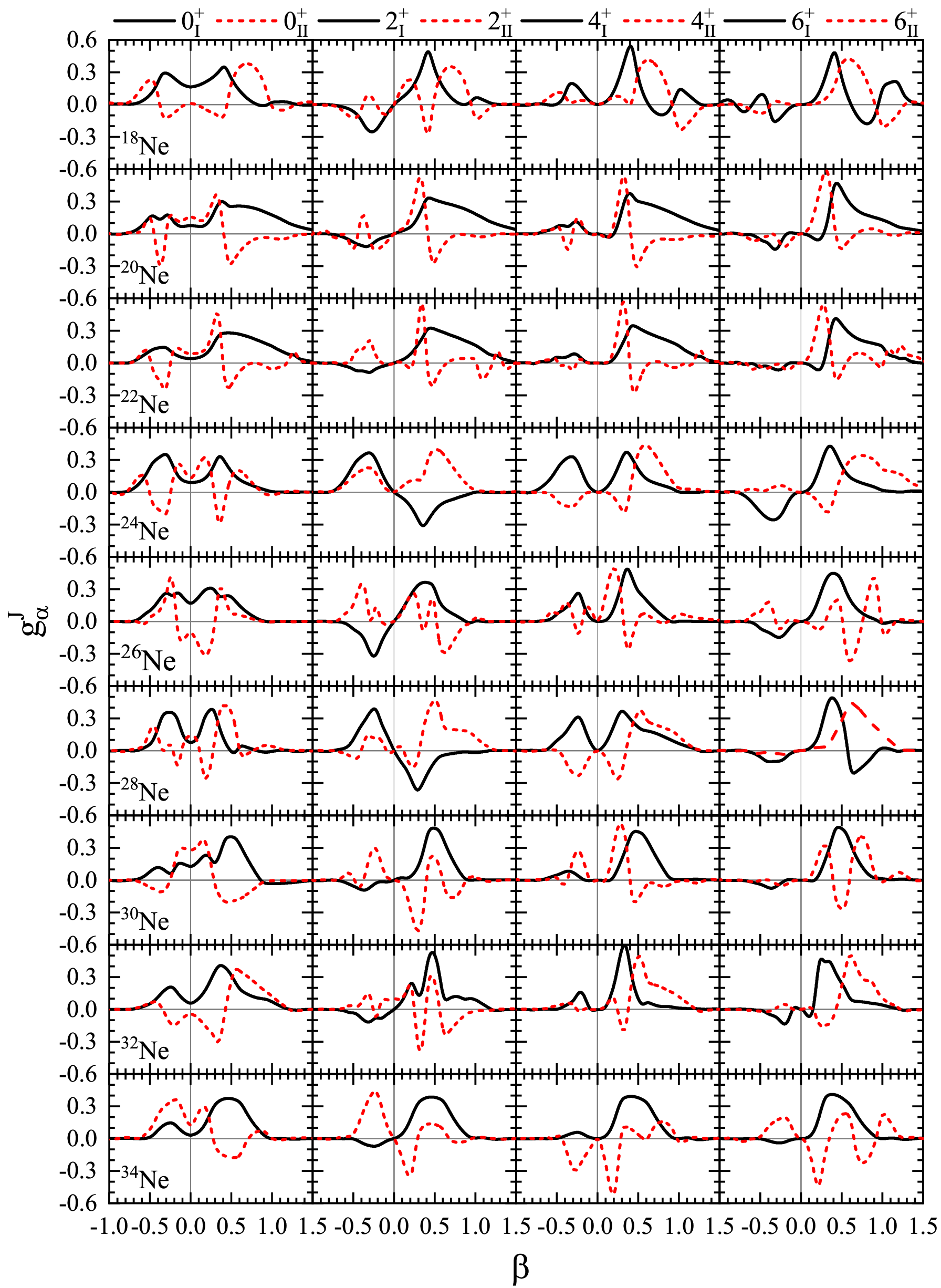}
	\caption{
		(Color online) Collective wave functions $g_{\alpha}^{J}$ for the low-lying states with 
		$J = 0, 2, 4$, and $6$ for $^{18\sim 34}$Ne as a function of the deformation of the mean-field states from which they are constructed. Where $g_{\rm I}^{J}$, $g_{\rm II}^{J}$ represents the wave functions of states that form the first and second band,
		respectively.
	}
	\label{f:gww}
\end{figure*}

\begin{figure*}[ht]
	\includegraphics[scale=0.5]{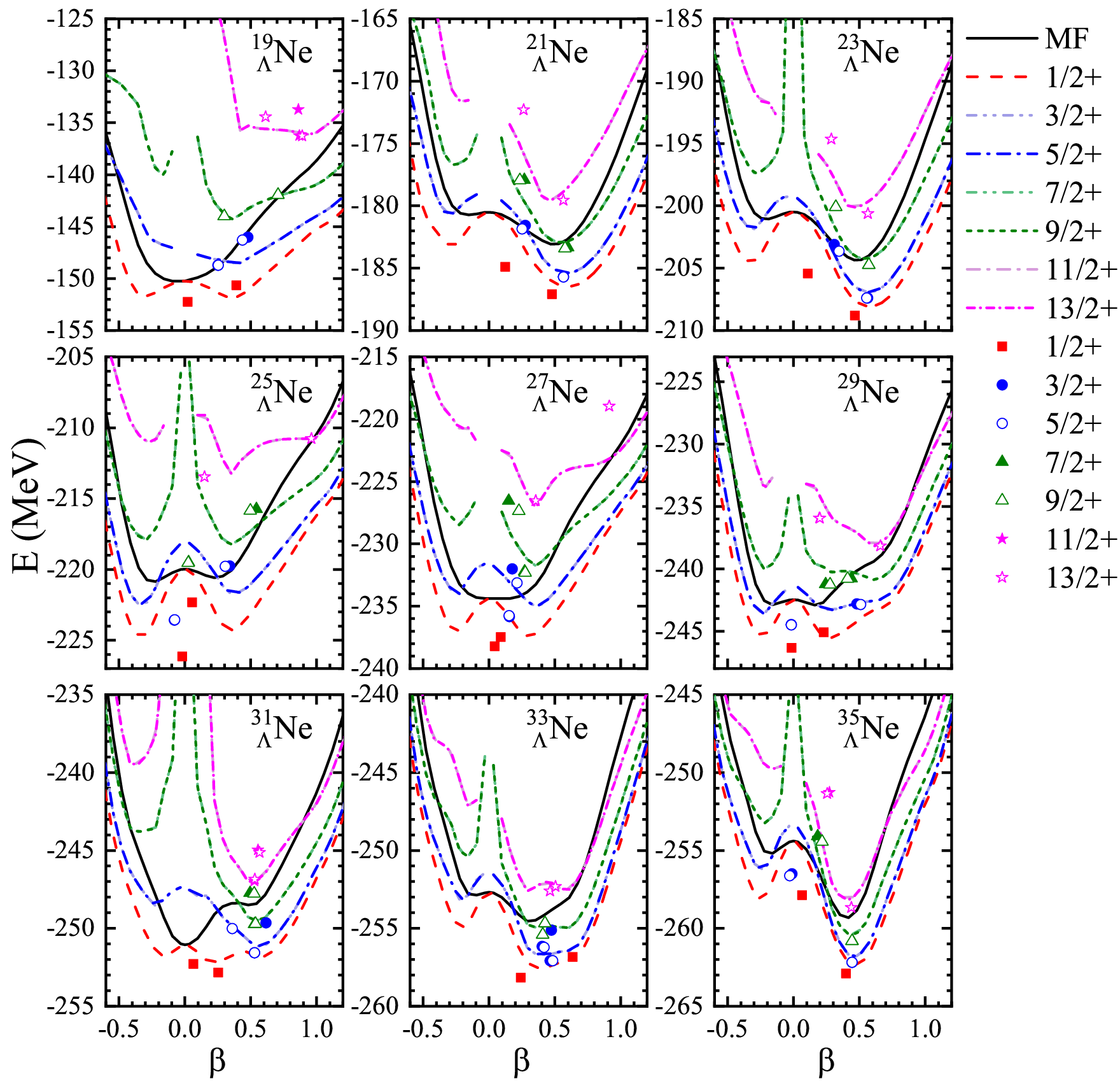}
	\caption{
		(Color online) Collective wave functions $g_{\alpha}^{J}$ for the low-lying states with 
		$J = 1/2,~3(5)/2,~7(9)/2$, and $11(13)/2$ for Ne-isotopes as a function of the deformation of the mean-field states from which they are constructed. 
	}
	\label{f:eps-amp-hyp}
\end{figure*}

\begin{figure*}[ht]
	\includegraphics[scale=0.5]{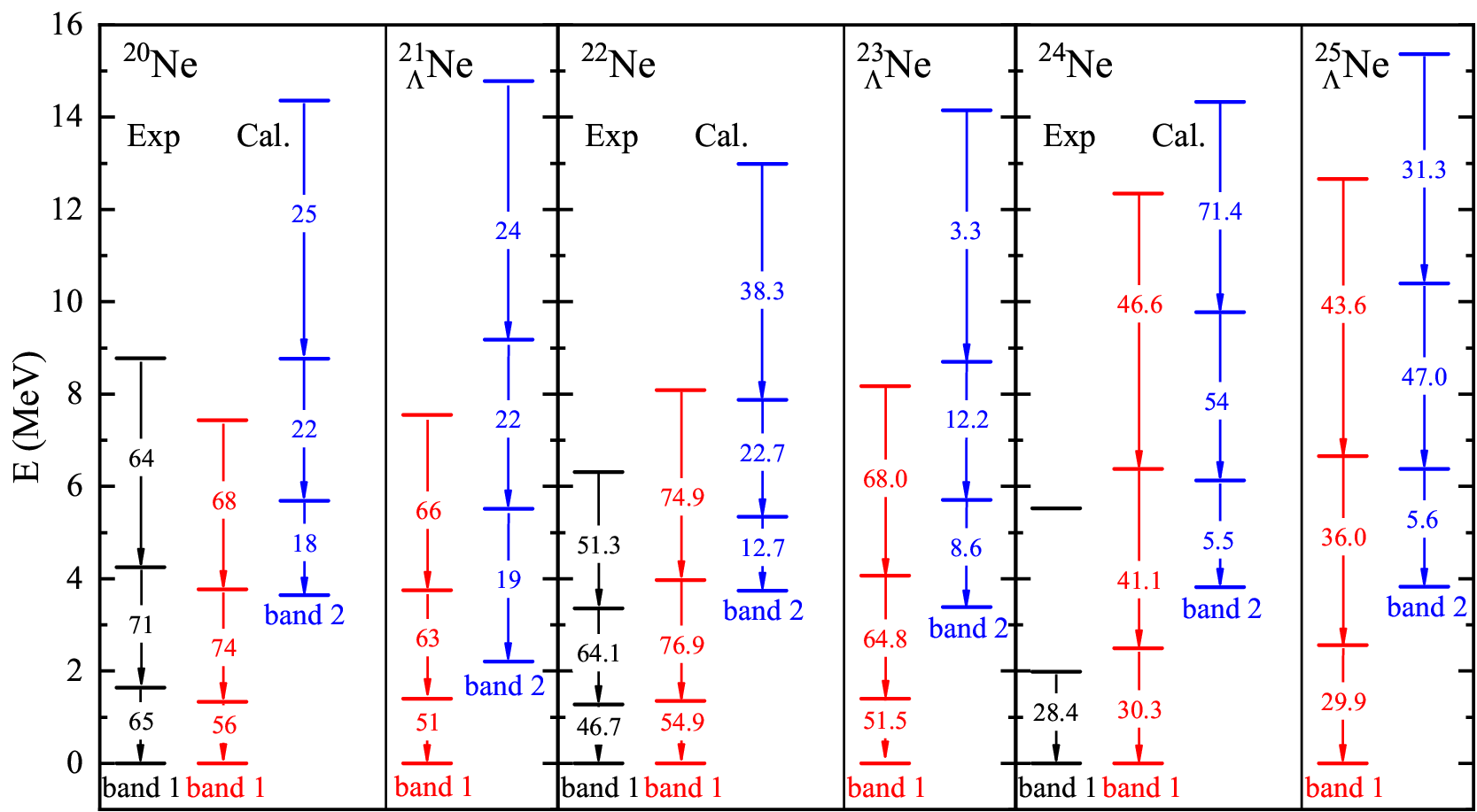}
	\caption{
		(Color online) 
		Same as Fig.\ref{f:level} but for comparion of hypernuclei and nuclei.
	}
	\label{f:level-hyp}
\end{figure*}

\begin{figure}[ht]
	\includegraphics[scale=0.3]{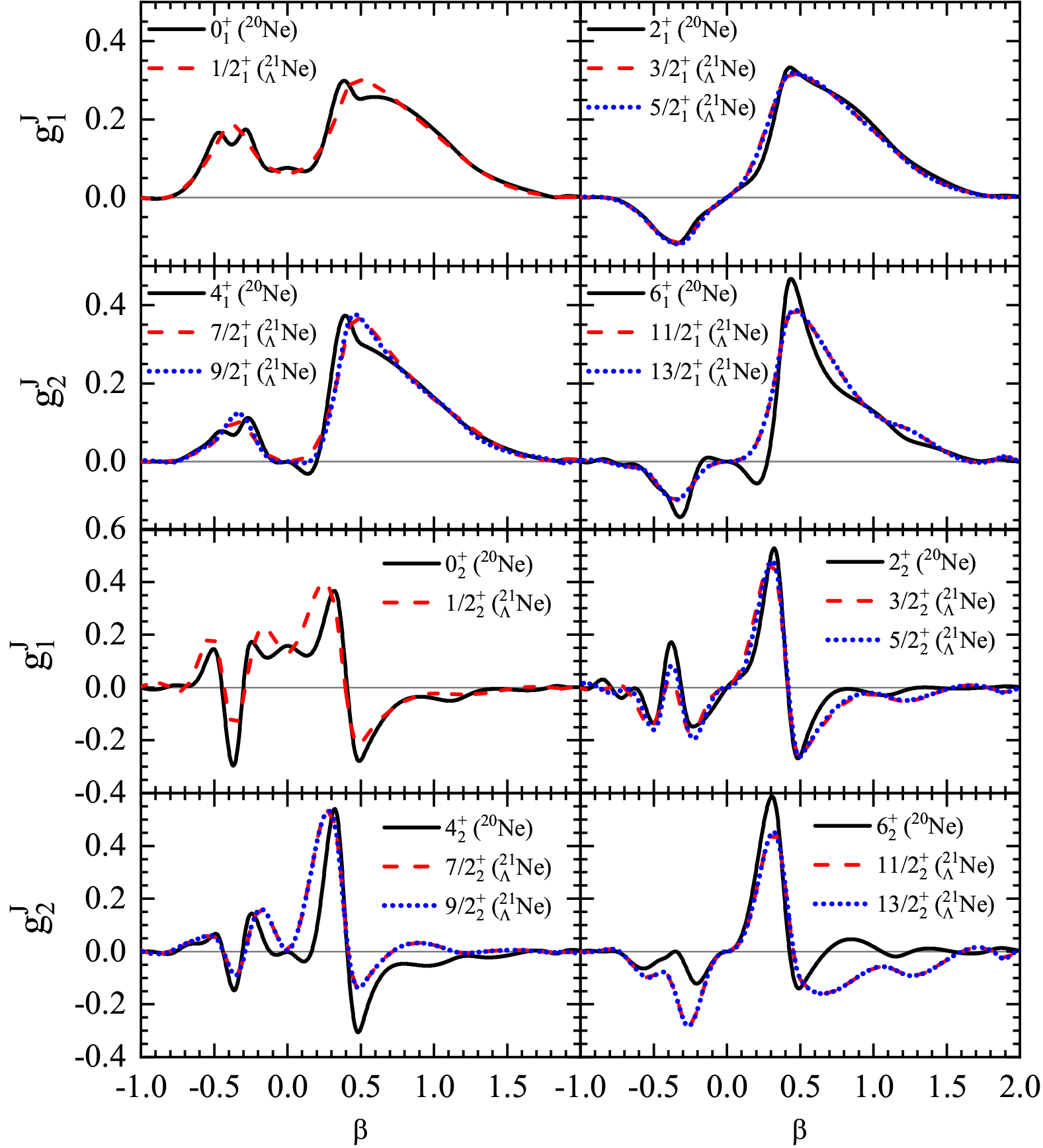}
	\caption{
		(Color online) Collective wave functions $g_{\alpha}^{J}$ for the low-lying states with 
		$J = 0, 2, 4$, and $6$ for $^{20}$Ne and $J = 1/2, 3/2, 5/2, 7/9, 11/2, 13/2$ 
		for $_{~\la}^{21}$Ne
		as a function of the deformation of the mean-field states from which they 
		are constructed. 
	}
	\label{f:Ne21gww1}
\end{figure}





\section{Results and discussion}
\label{s:results}
We first focus on the Ne-isotopes in the mean-field aspect, 
and then the AMP effect on shape coexistence and the low-lying 
spectrum for Ne-isotopes are discussed. 
This enables us to study the possibility of shape coexistence 
at different levels.
We will return to the more prospective case of Ne hyperisotopes at the end.

\subsection{Shape coexistence in Ne-isotopes}


\begin{table}
	\centering
	\caption{ 
		Ground-state deformations obtained from different calculations compared with the observed ones.
		Among them, $\beta_{\rm MF}$ and $\beta_{\rm AMP}$  indicate the minima of the mean-field PES and AMP PES, respectively. $\bar{\beta}$(Eq.~(\ref{e:bb})), $\overline{|\beta|}$ (Eq.~(\ref{e:bb2}))  are the average deformation of the ground state ($0^+$) given by configuration mixing of this current paper. It can indicate the signal of shape coexistence by comparing  $\bar{\beta}$(Eq.~(\ref{e:bb})) with  $\overline{|\beta|}$ (Eq.~(\ref{e:bb2})) . 
		$\bar{\beta}_{\rm AMD}$ is the average deformation of the ground state calculated by antisymmetric molecular dynamics, taken from Ref.~\cite{Sumi2012prc}.
		Because the experimental deformation are deduced from $B(E2)$ values, 
		the sign of deformation is not known, so the absolute value $|\beta_{\rm exp.}|$~\cite{NuDat} are 
		adopt here.
	}
	\renewcommand\arraystretch{1.5}
	\setlength{\tabcolsep}{1.5pt}
	\begin{ruledtabular}
		\begin{tabular}{cccccc|cc}
			& $|\beta_{\rm exp.}|$~\cite{NuDat} & $\beta_{\rm MF}$  & $\beta_{\rm AMP}$  
			&  $\bar{\beta}$ & $\overline{|\beta|}$
			&$\bar{\beta}_{\rm AMD}$~\cite{Sumi2012prc} & $\beta_{\rm HFB}$~\cite{hilaire2007epja,phynu}\\ 
			\hline 
			$^{18}$Ne & 0.68  & 0.0   & 0.36  & 0.10 & 0.31 & -    & 0.0 \\
			$^{20}$Ne & 0.72  & 0.53  & 0.59  & 0.52 & 0.63 & 0.46 & 0.50 \\
			$^{22}$Ne & 0.57  & 0.51  & 0.62  & 0.50 & 0.59 & 0.39 & 0.50 \\
			$^{24}$Ne & 0.41  & $-0.22$ & $-0.35$ & $-0.03$& 0.36 & $-0.25$& 0.25 \\
			$^{26}$Ne & 0.39  & 0.16  & 0.29  & 0.09 & 0.27 & 0.22 & 0.00 \\
			$^{28}$Ne & 0.36  & 0.16  & 0.29  & $-0.02$& 0.26& $-0.28$& 0.00 \\
			$^{30}$Ne & 0.49  & 0.0   & 0.16  & 0.44 & 0.48& 0.39 & 0.00 \\
			$^{32}$Ne &  -    & 0.35  & 0.35  & 0.31 & 0.39& 0.33 & 0.40 \\
			$^{34}$Ne &  -    & 0.42  & 0.48  & $-0.29$& 0.46& -    & - \\  
		\end{tabular}
	\end{ruledtabular}
	\label{t:beta}
\end{table}

\begin{table*}[htb]
	\caption{Transition rates $B(E2)$, Eq.~(\ref{e:BE2}), (in units of $e^2\text{fm}^4$) and the ratios $\Gamma_B$, Eq.~(\ref{e:Gamma}).  }
	\centering
	\renewcommand\arraystretch{1.5}
	\setlength{\tabcolsep}{1.5pt}
	\def\mr#1{\multirow{2}{*}{#1}}
	\begin{ruledtabular}
		\begin{tabular}{cccccccccccccccccccc}
			& \multicolumn{2}{c}{$_{\la}^{19}$Ne} & \multicolumn{2}{c}{$_{\la}^{21}$Ne} & \multicolumn{2}{c}{$_{\la}^{23}$Ne} 
			& \multicolumn{2}{c}{$_{\la}^{25}$Ne} & \multicolumn{2}{c}{$_{\la}^{27}$Ne} & \multicolumn{2}{c}{$_{\la}^{29}$Ne} 
			& \multicolumn{2}{c}{$_{\la}^{31}$Ne} & \multicolumn{2}{c}{$_{\la}^{33}$Ne} & \multicolumn{2}{c}{$_{\la}^{35}$Ne}\\
			$J_i \ra J_f$ &$B(E2)$ &$\Gamma_B$  & $B(E2)$ & $\bm{ \Gamma_B}$ &$B(E2)$ &$\bm{ \Gamma_B}$ &$B(E2)$ & $\bm{\Gamma_B}$ 
			&$B(E2)$ &$\Gamma_B$ &$B(E2)$ &$\Gamma_B$ &$B(E2)$  &$\Gamma_B$  &$B(E2)$ &$\Gamma_B$ &$B(E2)$ &$\bm{ \Gamma_B}$\\
			\hline
			$3/2_{\rm I}^+ \ra 1/2_{\rm I}^+$  &34.1 &0.88 & 51.2 & 0.92 &51.5 &0.94 &29.9 &0.99 &24.4 &0.84 &20.2 &0.78 &32.3 &0.63 &37.6 &0.95 &54.3 &0.95\\
			$5/2_{\rm I}^+ \ra 1/2_{\rm I}^+$  &33.8 &0.87 & 51.3 & 0.92 &51.6 &0.94 &29.8 &0.98 &26.9 &0.92 &20.2 &0.78 &33.8 &0.66 &36.2 &0.92 &54.4 &0.95\\
			$7/2_{\rm I}^+ \ra 3/2_{\rm I}^+$  &49.2 &1.03 & 62.7 & 0.85 &64.8 &0.84 &36.0 &0.88 &40.6 &0.93 &28.2 &0.88 &68.4 &0.87 &63.4 &3.62 &69.1 &0.86\\
			$9/2_{\rm I}^+ \ra 5/2_{\rm I}^+$  &54.3 &1.14 & 67.7 & 0.91 &71.7 &0.93 &40.2 &0.98 &43.6 &1.0  &29.5 &0.92 &76.6 &0.97 &70.2 &4.01 &76.8 &0.95\\
			$11/2_{\rm I}^+ \ra 7/2_{\rm I}^+$ &11.3 &0.17 & 65.6 & 0.96 &68.0 &0.91 &43.6 &0.94 &48.3 &1.01 &26.2 &1.53 &65.7 &0.79 &57.3 &1.39 &76.4 &0.92\\
			$13/2_{\rm I}^+ \ra 9/2_{\rm I}^+$ &24.5 &0.37 & 66.3 & 0.98 &70.6 &0.94 &45.4 &0.97 &47.1 &0.99 &24.8 &1.45 &53.6 &0.65 &59.6 &1.45 &79.2 &0.96\\
			\hline
			$J_i \ra J_f$ &$B(E2)$ &$\Gamma_B$  & $B(E2)$ & $ \Gamma_B$ &$B(E2)$ &$ \Gamma_B$ &$B(E2)$ & $\Gamma_B$ 
			&$B(E2)$ &$\Gamma_B$ &$B(E2)$ &$\Gamma_B$ &$B(E2)$  &$\Gamma_B$  &$B(E2)$ &$\Gamma_B$ &$B(E2)$ &$ \Gamma_B$\\
			$3/2_{\rm II}^+ \ra 1/2_{\rm II}^+$  &42.4 &0.99 & 18.7 & 1.04 &8.60 &0.68 &5.6  &1.02 &2.1  &0.36 &15.5 &0.66 &5.6  &0.64 &24.1 &0.41 &22.4 &1.10\\
			$5/2_{\rm II}^+ \ra 1/2_{\rm II}^+$  &46.9 &1.10 & 18.9 & 1.05 &8.90 &0.70 &4.7  &0.85 &11.6 &3.5  &16.4 &0.69 &2.7  &0.31 &12.1 &0.21 &21.6 &1.06\\
			$7/2_{\rm II}^+ \ra 3/2_{\rm II}^+$  &47.8 &0.78 & 21.9 & 0.99 &12.2 &0.54 &47.0 &0.87 &0.09 &0.06 &48.9 &0.67 &67.9 &1.90 &0.30 &0.01 &30.2 &1.61\\
			$9/2_{\rm II}^+ \ra 5/2_{\rm II}^+$  &46.0 &0.75 & 25.2 & 1.15 &27.8 &1.22 &55.5 &1.03 &28.1 &18.7 &51.8 &0.71 &64.5 &1.80 &1.6  &0.04 &31.7 &1.69\\
			$11/2_{\rm II}^+ \ra 7/2_{\rm II}^+$ &7.20 &0.07 & 24.4 & 0.98 &3.30 &0.09 &31.3 &0.44 &0.4  &0.06 &71.0 &1.11 &66.2 &6.13 &60.0 &0.92 &48.4 &1.30\\
			$13/2_{\rm II}^+ \ra 9/2_{\rm II}^+$ &5.10 &0.05 & 34.9 & 1.40 &3.40 &0.09 &30.0 &0.42 &0.7  &0.1  &67.7 &1.06 &58.0 &5.37 &62.3 &0.95 &49.8 &1.34\\
		\end{tabular}
	\end{ruledtabular}
	\label{t:E2}
\end{table*}

\begin{table*}[htb]
	\caption{
		The excitation energies $E$,
		rms charge radii $R_c$,
		and average deformations $\bbe$
		of the $0^+$, $2^+$, $4^+$ states of $^{20,22,24,34}$Ne and
		the $1/2^+$, $3/2^+$, $5/2^+$, $7/2^+$, $9/2^+$ states of $\la(1s)$ $_{~~~~~~~~~~~~~\la}^{21,23,25,35}$Ne.   }
	\renewcommand\arraystretch{1.5}
	\setlength{\tabcolsep}{1.5pt}
	\def\mr#1{\multirow{2}{*}{#1}}
	\begin{ruledtabular}
		\begin{tabular}{cccc|cccc}
			\multicolumn{4}{c|}{$^{20}$Ne} & \multicolumn{4}{c}{$_{~\la}^{21}$Ne}\\
			& $E$ [MeV] & $R_c$[fm] & $\bbe$  
			& $E$ [MeV] & $R_c$[fm] & $\bbe$ & \\
			\hline
	$0_{\rm I}^+$   &  0.0   & 2.89  & 0.52   & $1/2_{\rm I}^+$        & -15.58         & 2.88  & 0.48 \\
	$2_{\rm I}^+$   &  1.36  & 2.89  & 0.62   &$3(5)/2_{\rm I}^+$     & -14.19 (-14.19) & 2.88 (2.88)  & 0.56 (0.56) \\
	$4_{\rm I}^+$   &  3.77  & 2.89  & 0.60   &$7(9)/2_{\rm I}^+$     & -11.83 (-11.90) & 2.88 (2.87)  & 0.60 (0.58) \\
	$6_{\rm I}^+$   &  7.43  & 2.86  & 0.57   &$11(13)/2_{\rm I}^+$   & -8.04 (-8.04)   & 2.86 (2.86)  & 0.56 (0.56)  \\
			\hline
	$0_{\rm II}^+$   &  3.65  & 2.92  & 0.05  & $1/2_{\rm II}^+$     & -13.38          & 2.73  & 0.12 \\
	$2_{\rm II}^+$   &  5.69  & 2.92  & 0.18  & $3(5)/2_{\rm II}^+$  & -10.07 (-10.35) & 2.81 (2.79)  & 0.27 (0.25) \\
	$4_{\rm II}^+$   &  8.77  & 2.91  & 0.23  & $7(9)/2_{\rm II}^+$  & -6.41 (-6.45)   & 2.89 (2.88)  & 0.27 (0.23) \\
	$6_{\rm II}^+$   &  14.36 & 2.88  & 0.16  & $11(13)/2_{\rm II}^+$ & -0.80 (-0.80)  & 2.93 (2.93)  & 0.26 (0.26) \\
			\hline
			\multicolumn{4}{c|}{$^{22}$Ne} & \multicolumn{4}{c}{$_{~\la}^{23}$Ne}\\
			& $E$ [MeV] & $R_c$[fm] & $\bbe$ 
			& $E$ [MeV] & $R_c$[fm] & $\bbe$ & \\		
			\hline
	$0_{\rm I}^+$   &  0.0   & 2.96  & 0.50  & $1/2_{\rm I}^+$         & -16.33           & 2.94        & 0.47  \\
	$2_{\rm I}^+$   &  1.35  & 2.96  & 0.58  & $3(5)/2_{\rm I}^+$     & -14.93 (-14.93)  & 2.94 (2.94) & 0.56 (0.56) \\
	$4_{\rm I}^+$   &  3.97  & 2.95  & 0.59  & $7(9)/2_{\rm I}^+$     & -12.26 (-12.26)  & 2.93 (2.93) & 0.57 (0.57)\\
	$6_{\rm I}^+$   &  8.09  & 2.94  & 0.58  & $11(13)/2_{\rm I}^+$   & -8.16 (-8.16)    & 2.93 (2.92) & 0.56 (0.56) \\
			\hline
	$0_{\rm II}^+$   &  3.74  & 2.85  & 0.18 &   $1/2_{\rm II}^+$     & -12.94           & 2.81        & 0.11 \\
	$2_{\rm II}^+$   &  5.34  & 2.87  & 0.35 &   $3(5)/2_{\rm II}^+$ & -10.62 (-11.17)  & 2.83 (2.84) & 0.31 (0.34)\\
	$4_{\rm II}^+$   &  7.88  & 2.94  & 0.40 &   $7(9)/2_{\rm II}^+$ & -7.63 (-7.63)    & 2.88 (2.88) & 0.32 (0.32)\\
	$6_{\rm II}^+$   &  12.99 & 2.99  & 0.36 &   $11(13)/2_{\rm II}^+$ & -2.18 (-2.18)    & 2.94 (2.94) & 0.29 (0.29)\\	
			\hline
			\multicolumn{4}{c|}{$^{24}$Ne} & \multicolumn{4}{c}{$_{~\la}^{25}$Ne}\\
			& $E$ [MeV] & $R_c$[fm] & $\bbe$ 
			& $E$ [MeV] & $R_c$[fm] & $\bbe$ & \\		
			\hline
	$0_{\rm I}^+$   &  0.0   & 2.97  & -0.03  & $1/2_{\rm I}^+$         & -17.10          & 2.95         & -0.02 \\
	$2_{\rm I}^+$   &  2.49  & 2.97  & -0.09  & $3(5)/2_{\rm I}^+$      & -14.54 (-14.52)  & 2.95 (2.95)  & -0.08 (-0.08) \\
	$4_{\rm I}^+$   &  6.38  & 2.97  & -0.03  & $7(9)/2_{\rm I}^+$     & -10.44 (-10.49)  & 2.95 (2.95)  & 0.03 (0.03)\\
	$6_{\rm I}^+$   &  12.35 & 2.97  & 0.16   & $11(13)/2_{\rm I}^+$    & -4.44  (-4.44)   & 2.95 (2.95)  & 0.15 (0.15) \\
			\hline
	$0_{\rm II}^+$   &  3.81  & 2.93  & -0.09  & $1/2_{\rm II}^+$     & -13.28           & 2.92        & -0.05 \\
	$2_{\rm II}^+$   &  6.13  & 3.01  & 0.33   & $3(5)/2_{\rm II}^+$   & -10.72 (-10.74)  & 3.00 (2.99) & 0.34 (0.31) \\
	$4_{\rm II}^+$   &  9.77  & 3.05  & 0.54   & $7(9)/2_{\rm II}^+$   & -6.70  (-6.81)   & 3.03 (3.03) & 0.54 (0.50)\\
	$6_{\rm II}^+$   &  14.33 & 3.15  & 0.77   & $11(13)/2_{\rm II}^+$ & -1.73  (-1.73)   & 3.21 (3.21) & 0.96 (0.96)\\	
	\hline
				\multicolumn{4}{c|}{$^{34}$Ne} & \multicolumn{4}{c}{$_{~\la}^{35}$Ne}\\
	& $E$ [MeV] & $R_c$[fm] & $\bbe$ 
	& $E$ [MeV] & $R_c$[fm] & $\bbe$ & \\		
	\hline
	$0_{\rm I}^+$   &  0.0   & 3.42  & 0.41  & $1/2_{\rm I}^+$         & -18.62            & 3.40         & 0.40 \\
	$2_{\rm I}^+$   &  0.71  & 3.42  & 0.45  & $3(5)/2_{\rm I}^+$       & -17.89 (-17.89)   & 3.40 (3.40)  & 0.44 (0.44) \\
	$4_{\rm I}^+$   &  2.06  & 3.41  & 0.45  & $7(9)/2_{\rm I}^+$      & -16.52 (-16.53)   & 3.40 (3.40)  & 0.45 (0.44)\\
	$6_{\rm I}^+$   &  4.33  & 3.40  & 0.45  & $11(13)/2_{\rm I}^+$    & -14.38  (-14.38)  & 3.38 (3.38)  & 0.44 (0.44) \\
	\hline
	$0_{\rm II}^+$   &  5.49  & 3.35  & 0.02 &   $1/2_{\rm II}^+$      & -13.59            & 3.33        & 0.06 \\
	$2_{\rm II}^+$   &  6.54  & 3.35  & 0.06 &   $3(5)/2_{\rm II}^+$   & -12.23 (-12.34)   & 3.35 (3.34) & 0.02 (0.03) \\
	$4_{\rm II}^+$   &  8.81  & 3.36  & 0.14 &   $7(9)/2_{\rm II}^+$   & -9.81  (-10.16)   & 3.37 (3.37) & 0.18 (0.22)\\
	$6_{\rm II}^+$   &  11.81 & 3.50  & 0.38 &   $11(13)/2_{\rm II}^+$ & -6.98  (-7.07)    & 3.40 (3.39) & 0.27 (0.25)\\
		\end{tabular}
	\end{ruledtabular}
	\label{t:r}
\end{table*}

Table~\ref{t:beta} shows quadrupole deformations of the Ne-isotopes in this current work, compared to the ones from experiments and the ones given by other models(AMD and HFB).  
It can be seen that, except for $^{24}$Ne,  $^{26}$Ne, and $^{28}$Ne, the deformation obtained in this work, especially the average deformation  $\bbe$  of the ground state, is reasonably consistent with the experimental one, which is closer to the experimental value than that calculated by AMD. 
In addition, the $\overline{|\be|}$ in Eq.~(\ref{e:bb2}) matches 
the $|\beta_{\rm exp.}|$ better than $\bbe$ in Eq.~(\ref{e:bb}), 
especially for $^{24,26,28}$Ne. For example, the $\bbe$ of $^{24}$Ne is $-0.03$, 
while the $\overline{|\be|}$ is $0.36$, which is closer to $0.41$.
Because the deformation of the two dominant configurations is close but with opposite signs, 
the degree of deformation is cancelled out and not truthfully reflected in $\bbe$ in this nucleus, 
which is also reflected in the collective wave functions. As shown in Fig.~\ref{f:gww}, 
one can see a strong cancellation between prolate and oblate contributions. 
This is a sign of configuration mixing of two minima in Fig.~\ref{f:eps-amp}, 
i.e., a kind of shape coexistence but not the spherical shape. This will be discussed in detail later.
This fully demonstrates that the results of this work are reliable to a certain extent.

The mean field PESs (black solid line) are shown in Fig.~\ref{f:eps-amp}. 
While the PES of $^{18}$Ne has a spherical minimum, both $^{20}$Ne and $^{22}$Ne are 
prolate deformed in their ground states. In $^{20}$Ne, the prolate ground state corresponds 
to $\beta = 0.53$, and an oblate local minimum also appears at $\beta = -0.16$ with 
an excitation energy of $2.71$ MeV. In the case of $^{22}$Ne, the ground state 
corresponds to $\beta = 0.51$, and another local minimum is found at $\beta = -0.24$ 
with an excitation energy of $3.54$ MeV. The nucleus $^{24}$Ne is a clear example of 
very strong shape coexistence in the considered isotopic chain since, 
while the oblate ground state is located at $\beta = -0.22$, a prolate isomeric state 
is also found at $\beta = 0.22$ with an excitation energy with respect to the oblate 
ground state of $300$ keV, which roughly agrees with the $77$ keV obtained by Gogny mean field~\cite{Rodriguez2003}.

On the other hand, the nuclei $^{26, 30}$Ne show spherical ground states 
indicating that the  $N =16$ subshell closure~\cite{Obertelli2005prc} 
and N$=20$ shell closure are preserved at the mean-field level. 
The MFPES of both $^{26}$Ne and $^{30}$Ne are particularly flat around their spherical ground states. 
In the case of $^{28}$Ne, the ground state corresponds to $\beta = 0.16$, 
and another local minimum is found at $\beta = -0.16$ with an excitation energy of $30$ keV 
higher than the ground state, indicating the existence of shape coexistence in this nucleus as well.
In the nucleus $^{30}$Ne we obtain a prolate shoulder at $\beta=0.42$ at 
an excitation energy of $2.32$~MeV with respect to the spherical ground state. 
In the drip line systems $^{32}$Ne and $^{34}$Ne, prolate deformed 
ground states are found. 
The ground states have $\beta=0.35$ and $\beta= 0.42$, respectively. 
In addition, an oblate isomeric state is found in $^{32}$Ne/ $^{34}$Ne at $\beta=0.16/0.12$ 
with an excitation energy of $1.89/2.39$ MeV with respect to the prolate ground state.

Before considering the full AMP-GCM, it is instructive to look into the angular momentum projected potential energy surfaces (AMPPES) defined as~Eq.~(\ref{e:ejk}).
The corresponding MF and AMP energy landscapes are also included for comparison.
For details on the missing points in the $I = 2, 4$ and $6$ curves refer to~\cite{Rodriguez2002NPA}. 
The most remarkable fact about Fig.~\ref{f:eps-amp} is how strongly the restoration of the rotational 
symmetry modifies the mean-field picture of the $I = 0$ configurations.
 For most isotopes, the energy barrier between the two minima is enhanced due to 
the restoration of rotational symmetry.

The prolate minimum is, with the exception of $^{24}$Ne, the absolute minimum 
in all the isotopes considered, which is consistent with Gogny results~\cite{Rodriguez2003}. 
The orbital responsible for such an oblate minimum is the neutron 
$1d_{5/2}$ orbital which becomes fully occupied in $^{24}$Ne and favors oblate deformations.
With increasing spin values either the energy difference between 
the prolate and oblate minima increases or the oblate minimum is washed out.
In addition, shape coexistence is expected in the nuclei $^{24}$Ne,
$^{26}$Ne, $^{28}$Ne and $^{30}$Ne as their  $I^{\pi}=0^{+}$ prolate and oblate minima 
are very close in energy($446,384, 303$ and $368$ keV, respectively). 
These minima are separated by barriers which are $5.0,3.1, 3.3$ and $1.1$ MeV high, respectively.

The AMPPESs show the phenomenon of shape coexistence for some nuclei and/or some spin values, 
and therefore configuration mixing has to be considered in order to gain a better understanding 
of the structure of these states. In Fig.~\ref{f:eps-amp}, the lowest GCM energy levels 
for $J = 0$ up to $6$ are given. The collective levels are plotted at the mean deformation $\bar{\beta}$ 
of the mean-field states from which they are constructed, defined as in Eq~.(\ref{e:bb}), 
which provides in many cases an intuitive picture of the band structure in a nucleus.

The prominent feature is that configuration mixing decreases the average deformation $\bbe$ of 
the ground states with respect to the minimum of the AMPPES. 
The ground states of the nuclei $^{24}$Ne, $^{26}$Ne and $^{28}$Ne become spherical, 
becaues the prolated and oblated minimum are cancelled out.
The other nuclei remain well deformed in their ground states and develop 
a rather well-defined rotational band up to the maximum spin considered for $\alpha=1$. 
In addition, a well-defined rotational band is obtained for $J\geq 2$ in $^{20}$Ne,
$^{22}$Ne, and $^{24}$Ne. 
On the other hand, the excited states ($\alpha=2$) only show a rotational band 
pattern for those nuclei well deformed in their ground state.


In order to provide a more detailed description of the rotational band 
mentioned above, we present a comparison between the ground-state excitation energy 
of the rotational band~(almost all the ground state bands of Ne isotopes are composed of
states with $\alpha=1$, except for $^{28}$Ne) 
and the experimental values in Fig.~\ref{f:level}, 
In the same figure, the theoretical excitation energies and the values of $B(E2)$ for 
the possible second band are also given.


Based on the comparison of the energy of each excited state and $B(E2)$ values between them 
in the ground state band shown in Fig.~\ref{f:level} with the experimental values, 
it can be said that our calculation reasonably provides a rotational band 
that is consistent with the experimental values for every Ne isotope except for $^{18}$Ne. 
In addition, the excitation energy levels and predicted $B(E2)$ values in the second 
band of these nuclei are also given in Fig.~\ref{f:level}.

However, unlike the structure of the first band that is clearly related to rotation, 
the second band exhibits structural characteristics of vibration bands, 
some of which are more like $\beta$-vibration bands, 
which we will discuss in detail later in conjunction with wave functions.
 For $^{20 \sim 24,34}$Ne, the gap between the 
energy level in the band established based on the second $0^+$ is
roughly close to that of the ground state band.
The wave function of these states of the second band in the nuclei 
exhibit positive and negative phase oscillations as shown in Fig~\ref{f:gww}.
The average deformation $\bar{\beta}$ of the states in the second band increases 
with the increase of angular momentum $J$, which reflects 
the fluctuations of the excited states on the 
shape parameter $\beta$ and also reflects the increasingly 
important contribution of the prolated configuration.
Given this, it is highly likely a $\beta$ vibration band caused 
by collective shape fluctuations.

For the second band of $^{28,30}$Ne, the spacing of energy levels within the band 
are roughly equal, exhibiting characteristics of vibrational bands (harmonic like spectrum~\cite{Peter1980}). 
As is well known, vibrational band is generated based on spherical nucleus. 
Since the $\bar{\beta}$ of the states with $J\le 4$  in the 
second band of $^{30}$Ne are close to $0$,  
it can be considered that the part of second band is a vibrational band. 
On the other hand, the first and second bands of $^{32}$Ne are more like rotational bands 
built on different cores because the energy levels within each band have the characteristics 
of rotational bands and the average deformation of each state within a band is roughly equal.



Fig.~\ref{f:gww}, however, illustrates the limits of the meaning of $\bar{\beta}$, 
showing the collective wave functions of the states with $J = 0, 2, 4$, and $6$. 
All low-lying states, i.e., $0^{+}, 2^{+}, 4^{+}$, and $6^{+}$ of the ground band result 
from mixing between prolate and oblate reference states. 
Especially for nuclei with spherical or weakly deformed ground states, such as $^{18,24,26,28,30}$Ne, 
their ground states exhibit a strong mixing of prolate and oblate reference states, 
which leads to spherical ground states on average. However, as the angular momentum $J$ increases, 
the contribution of prolate reference states increases. For nuclei with well-deformed ground states, 
the dominance of prolate deformations is evident. Returning to Fig.~\ref{f:eps-amp}, 
the values of $\bar{\beta}$ for the states in the ground band reflect this dominance. 
For example, in $^{20}$Ne, the $\bar{\beta}$ of each state in the ground state band ($0_1^+,~2_1^+,~4_1^+,~6_1^+$) is approximately equal to $0.5$, which is consistent with the $\beta$ of the energy minimum on the prolate side of their projected energy curves.

For states in the second band, the situation is similar to that of the ground band.
The very small value of $\bar{\beta}$ does not mean that 
this state is nearly spherical, but rather, that the weights of prolate and oblate shapes are nearly equal.
For higher $J$ values, the mixing between oblate and prolate
configurations are less pronounced as shown in Fig.~\ref{f:gww}, and the value of $\bar{\beta}$ better
represents the structure of the states.
For example, in $^{20}$Ne,
$\bar{\beta}$ of the $2_2^+,~4_2^+$, and $6_2^+$ excited states shown in 
Fig.~\ref{f:eps-amp} corresponds very well to the deformation values 
of the main components of these states shown in Fig.~\ref{f:gww}, approximately $0.3$.

\subsection{Impurity effects of $\la_{s}$ on low-lying states in Ne isotopes}


Next, let's focus on the influence of an $s$-state $\la$ hyperon on 
the structure of low-lying states in nuclei.
Firstly, by comparing the Fig.~\ref{f:eps-amp}, in which the MFPES and AMPPES, 
as well as the low-lying states with their average deformation $\bar{\beta}$ are included, 
with Fig.~\ref{f:eps-amp-hyp}, it can be seen that the influence of $s$-state $\la$  on the structure 
of low-lying states of the first band is not so significant, except for $_{~~~~\la}^{31,33}$Ne.
This is due to the fact that the $s$-orbit $\la$ is spherically distributed (or mildly deformed) 
and thus does not change the shape of the nuclear core dramatically.
This is also demonstrated in Fig.~\ref{f:Ne21gww1}, $_{~\la}^{21}$Ne for example,
by the weights of the natural states in the collective subspace, Eq.~(\ref{e:g}).
Again the weights of the $1/2_1^+$($3/2_1^+\&5/2_1^+$, $7/2_1^+\&9/2_1^+$, $11/2_1^+\&13/2_1^+$) states
are similar to those of the corresponding $0^+$($2^+$, $4^+$, $6^+$) states. 
However, the addition of $s$-state $\la$ has caused significant changes to many low-lying
	states, especially for $_{~~~~\la}^{31,33}$Ne, where the band that has a 
	rotational-like structure in their two core nuclei has been disrupted. 
\textcolor{blue} This is somewhat different from the shrinkage effect of $\la$ on the ground state~\cite{Cui2015prc,Xue2024prc}.

In Table~\ref{t:E2}, we list the $B(E2)$ between states
within the ground and second bands of Ne-isotopes,
and of their corresponding hypernuclei.
Due to the splitting of angular momentum into $J\pm1/2$,
each of the $B(E2)$ values of the core nucleus
has two counterparts in hypernuclei,
which are both listed.
Unlike the shrinkage effect of $\la_s$ discovered previously 
on low-lying states~\cite{Cui2017prc,Xue2024prc},
where the addition of one $\la_s$ enhances or reduces the $B(E2)$ in 
the ground or second band.
This expansion and shrinkage effect are characterized by~\cite{Cui2015prc},
\begin{equation}\label{e:Gamma}
\Gamma_B=\frac{B\left(E2,J_i^+\to J_f^+;{}_{~~\la}^{A+1}\mathrm{Ne}\right)}{B\left(E2,J_i^+\to J_f^+;{}_{}^{A}\mathrm{Ne}\right)},
\end{equation}
also listed in Table~\ref{t:E2}.


In general, the $B(E2)$ are proportional to $R_c^4$ and to $\be^2$~\cite{Cui2017prc}.
Both the shrinkage of the nuclear size indicated by $R_c$ 
and the reduction of the quadrupole deformation $\bar\be$,
thus contribute to the overall reduction of $B(E2)$~\cite{Cui2017prc,Xue2024prc}.
As shown in Table~\ref{t:E2}, $\Lambda_s$ slightly reduces the $B (E2)$ 
within the ground band of $^{21 \sim 27, 35}$Ne. 
The ratios($\Gamma_{B}$) of $B(E2)$ between the states of  hypernuclei to $B(E2)$ 
between the states of  nuclei core are approximately $1$~(highlighted in bold in Table~\ref{t:E2}), 
while the ground bands of other isotopes are obviously influenced by $\la_s$.
It means that the deformations of the cores of $^{21 \sim 27, 35}$Ne 
are relatively stable compared to those of other hyperisotopes, 
and an $\la_s$ is not enough to change it.

    In addition, by comparing the $B(E2)$ within the ground band of the isotopes, 
	it was found that the influence of $\la$ on $_{~\Lambda}^{21}$Ne is greater than that 
	on $_{~\Lambda}^{23}$Ne and $_{~\Lambda}^{25}$Ne.
	It should be explained from the following: Fig.~\ref{f:eps-amp-hyp} shows 
	the potential energy surfaces of $_{~\Lambda}^{21}$Ne, $_{~\Lambda}^{23}$Ne, and $_{~\Lambda}^{25}$Ne. 
	We can clearly see the differences in the potential energy surfaces among these three isotopes: 
	the depths of the valleys formed on both sides of the oblate and prolate shapes for $_{~\Lambda}^{21}$Ne 
	are about $1.5$~MeV and $7$~MeV, respectively; for $_{~\Lambda}^{23}$Ne, 
	they are about $2.5$~MeV and $8$~MeV, respectively; and for $_{~\Lambda}^{25}$Ne, 
	they are about $5$~MeV and $4$~MeV. Taking $_{~\Lambda}^{21}$Ne as an example, 
	according to the wave functions shown in Fig.~\ref{f:Ne21gww1}, 
	the reference states in the collective space contribute more to 
	the ground band from the states within the two deeper valleys compared to states 
	at other deformations, hence deeper valleys may lead to more stable collective states. 
	Therefore, the ground band of $_{~\Lambda}^{21}$Ne, which is mainly composed of reference states 
	from the two valleys with depths of about $1.5$~MeV and $7$~MeV, is less stable compared to two others, 
	making it more susceptible to changes induced by the addition of the $\Lambda$ hyperon.

Further, the situation becomes more complicated for states in the second band of 
almost of Ne isotopes, 
where some $B(E2)$ increase or decrease. 
Such variations are also reflected in the $\bar{\beta}$ and rms radii. 
As shown in Table~\ref{t:r}, for some states, the average deformation $\bar{\beta}$ 
does not decrease due to the inclusion of $\Lambda_s$, 
but rather tends towards prolate deformation, while the radius decreases.

To investigate the impurity effect of $\la_{1s}$ on low-lying states,
the comparion of low-lying spectra of hypernuclei and nuclei are shown.
In Fig.~\ref{f:level-hyp}, the level structures of the second bands of $^{20}$Ne and $_{~\Lambda}^{21}$Ne, 
$^{22}$Ne and $_{~\Lambda}^{23}$Ne, 
and $^{24}$Ne and $_{~\Lambda}^{25}$Ne are given, and it is found that the addition of $\Lambda_{s}$ 
appears to affect the excitation modes of these bands. 
As shown in Fig.~\ref{f:level-hyp}, this phenomenon is particularly evident in $^{20}$Ne and $_{~\Lambda}^{21}$Ne, 
and is also reflected in $^{22}$Ne and $_{~\Lambda}^{23}$Ne, $^{24}$Ne and $_{~\Lambda}^{25}$Ne, 
where structures resembling $\beta$ vibration transition to 
vibration modes with equal energy gaps. 
This can be seen from the wavefunctions of $^{20}$Ne and $_{~\Lambda}^{21}$Ne given in the lower four subplots in Fig.~\ref{f:Ne21gww1}, 
where the addition of $\Lambda_s$ reduces the phase oscillation amplitude of 
the collective wave functions of states in the second band of $^{20}$Ne.

In Fig.~\ref{f:level-hyp}, the level structures of the second bands of $^{20}$Ne 
and $_{~\Lambda}^{21}$Ne, $^{22}$Ne and $_{~\Lambda}^{23}$Ne, and $^{24}$Ne and $_{~\Lambda}^{25}$Ne are given, 
and it is found that the addition of $\Lambda_{s}$ appears to affect the excitation modes of these bands. 
As shown in Fig.~\ref{f:level-hyp}, this phenomenon is particularly evident in $^{20}$Ne and $_{~\Lambda}^{21}$Ne, and is also reflected in $^{22}$Ne and $_{~\Lambda}^{23}$Ne, $^{24}$Ne and $_{~\Lambda}^{25}$Ne, where structures resembling $\beta$ vibration transition to vibration modes with equal energy gaps. This can be seen from the wave functions of $^{20}$Ne and $_{~\Lambda}^{21}$Ne given in the lower four subplots in Fig.~\ref{f:Ne21gww1}, where the addition of $\Lambda_s$ reduces the phase oscillation amplitude of the collective wave functions of states in the second band of $^{20}$Ne.

\section{Summary}
\label{s:summary}
In summary, based on the results of the beyond mean filed SHF approach, 
the shape coexistence of Ne isotopes is discussed at the mean field level 
and the beyond mean filed level, respectively. 
Then we studied the impurity effect of $\la$ on the low-lying spectra of these nuclei.
The results of the mean field indicate that there is shape coexistence in 
the two isotopes $^{24}$Ne and $^{28}$Ne, due to the presence of two minima 
on the potential energy surface with similar energy but completely different shapes. 
The angular momentum projection provides additional shape coexistence nuclei: $^{26}$Ne and $^{30}$Ne. 
The results of GCM indicate that the ground states of $^{24}$Ne, $^{26}$Ne, and $^{28}$Ne 
are not truly spherical, but are a mixture of prolate and oblate configurations, 
which is an obvious sign of shape coexistence.
In addition, well established rotational bands based on deformed ground states and $\beta$ vibrational bands, 
whose collective wave functions exhibit positive and negative phase oscillations 
with similar structures to rotational bands, were found in the isotopes $^{20-24,34}$Ne.

Next, we investigated the impurity effect of the $s$-state $\Lambda$ on the 
band structure of Ne isotopes with coexisting shapes. 
We found that the $s$-state $\Lambda$ has a shrinkage effect on the states in the ground band, similar to its effect on the ground state in the mean-field. 
However, more dramatic is the influence of the $\Lambda_s$ on the second band, which is quite unusual, 
as it seems to change the excitation mode of this band. The addition of the $\Lambda_s$ results in an equidistant orientation within the band, shifting it from a $\beta$ vibration band to a vibration band limit.

\section*{Acknowledgements}

This work was supported by the National Natural Science Foundation of China
under Grant Nos.~12175071, 12205103 and 11905165.

\newcommand{\epja}{EPJA}
\newcommand{\npa}{Nucl. Phys. A}
\newcommand{\nphysa}{Nucl. Phys. A}
\newcommand{\ppnp}{Prog. Part. Nucl. Phys.}
\newcommand{\ptp}{Prog. Theor. Phys.}
\newcommand{\ptep}{Prog. Theor. Exp. Phys.}
\bibliographystyle{apsrev4-1}
\bibliography{coexistence}

\begin{thebibliography}{58}
\expandafter\ifx\csname natexlab\endcsname\relax\def\natexlab#1{#1}\fi
\expandafter\ifx\csname bibnamefont\endcsname\relax
  \def\bibnamefont#1{#1}\fi
\expandafter\ifx\csname bibfnamefont\endcsname\relax
  \def\bibfnamefont#1{#1}\fi
\expandafter\ifx\csname citenamefont\endcsname\relax
  \def\citenamefont#1{#1}\fi
\expandafter\ifx\csname url\endcsname\relax
  \def\url#1{\texttt{#1}}\fi
\expandafter\ifx\csname urlprefix\endcsname\relax\def\urlprefix{URL }\fi
\providecommand{\bibinfo}[2]{#2}
\providecommand{\eprint}[2][]{\url{#2}}

\bibitem[{\citenamefont{Garrett et~al.}(2019)\citenamefont{Garrett,
  Rodr\'{\i}guez, Varela, Green, Bangay, Finlay, Austin, Ball, Bandyopadhyay,
  Bildstein et~al.}}]{Garrett2019PRL}
\bibinfo{author}{\bibfnamefont{P.~E.} \bibnamefont{Garrett}},
  \bibinfo{author}{\bibfnamefont{T.~R.} \bibnamefont{Rodr\'{\i}guez}},
  \bibinfo{author}{\bibfnamefont{A.~D.} \bibnamefont{Varela}},
  \bibinfo{author}{\bibfnamefont{K.~L.} \bibnamefont{Green}},
  \bibinfo{author}{\bibfnamefont{J.}~\bibnamefont{Bangay}},
  \bibinfo{author}{\bibfnamefont{A.}~\bibnamefont{Finlay}},
  \bibinfo{author}{\bibfnamefont{R.~A.~E.} \bibnamefont{Austin}},
  \bibinfo{author}{\bibfnamefont{G.~C.} \bibnamefont{Ball}},
  \bibinfo{author}{\bibfnamefont{D.~S.} \bibnamefont{Bandyopadhyay}},
  \bibinfo{author}{\bibfnamefont{V.}~\bibnamefont{Bildstein}},
  \bibnamefont{et~al.}, \bibinfo{journal}{Phys. Rev. Lett.}
  \textbf{\bibinfo{volume}{123}}, \bibinfo{pages}{142502}
  (\bibinfo{year}{2019}).

\bibitem[{\citenamefont{Aberg et~al.}(1990)\citenamefont{Aberg, Flocard, and
  Nazarewicz}}]{Aberg1990}
\bibinfo{author}{\bibfnamefont{S.}~\bibnamefont{Aberg}},
  \bibinfo{author}{\bibfnamefont{H.}~\bibnamefont{Flocard}}, \bibnamefont{and}
  \bibinfo{author}{\bibfnamefont{W.}~\bibnamefont{Nazarewicz}},
  \bibinfo{journal}{Annual Review of Nuclear and Particle Science}
  \textbf{\bibinfo{volume}{40}}, \bibinfo{pages}{439} (\bibinfo{year}{1990}).

\bibitem[{\citenamefont{Garrett et~al.}(2022)\citenamefont{Garrett, Zielińska,
  and Clément}}]{Garrett2022PLB}
\bibinfo{author}{\bibfnamefont{P.~E.} \bibnamefont{Garrett}},
  \bibinfo{author}{\bibfnamefont{M.}~\bibnamefont{Zielińska}},
  \bibnamefont{and} \bibinfo{author}{\bibfnamefont{E.}~\bibnamefont{Clément}},
  \bibinfo{journal}{Progress in Particle and Nuclear Physics}
  \textbf{\bibinfo{volume}{124}}, \bibinfo{pages}{103931}
  (\bibinfo{year}{2022}), ISSN \bibinfo{issn}{0146-6410}.

\bibitem[{\citenamefont{Heyde and Wood}(2011)}]{Heyde2011RMP}
\bibinfo{author}{\bibfnamefont{K.}~\bibnamefont{Heyde}} \bibnamefont{and}
  \bibinfo{author}{\bibfnamefont{J.~L.} \bibnamefont{Wood}},
  \bibinfo{journal}{Rev. Mod. Phys.} \textbf{\bibinfo{volume}{83}},
  \bibinfo{pages}{1467} (\bibinfo{year}{2011}).

\bibitem[{\citenamefont{Heyde et~al.}(1983)\citenamefont{Heyde, {Van Isacker},
  Waroquier, Wood, and Meyer}}]{Heyde1983PR}
\bibinfo{author}{\bibfnamefont{K.}~\bibnamefont{Heyde}},
  \bibinfo{author}{\bibfnamefont{P.}~\bibnamefont{{Van Isacker}}},
  \bibinfo{author}{\bibfnamefont{M.}~\bibnamefont{Waroquier}},
  \bibinfo{author}{\bibfnamefont{J.}~\bibnamefont{Wood}}, \bibnamefont{and}
  \bibinfo{author}{\bibfnamefont{R.}~\bibnamefont{Meyer}},
  \bibinfo{journal}{Physics Reports} \textbf{\bibinfo{volume}{102}},
  \bibinfo{pages}{291} (\bibinfo{year}{1983}), ISSN \bibinfo{issn}{0370-1573}.

\bibitem[{\citenamefont{Bonn et~al.}(1972)\citenamefont{Bonn, Huber, Kluge,
  Kugler, and Otten}}]{Bonn1972PLB}
\bibinfo{author}{\bibfnamefont{J.}~\bibnamefont{Bonn}},
  \bibinfo{author}{\bibfnamefont{G.}~\bibnamefont{Huber}},
  \bibinfo{author}{\bibfnamefont{H.-J.} \bibnamefont{Kluge}},
  \bibinfo{author}{\bibfnamefont{L.}~\bibnamefont{Kugler}}, \bibnamefont{and}
  \bibinfo{author}{\bibfnamefont{E.}~\bibnamefont{Otten}},
  \bibinfo{journal}{Physics Letters B} \textbf{\bibinfo{volume}{38}},
  \bibinfo{pages}{308} (\bibinfo{year}{1972}), ISSN \bibinfo{issn}{0370-2693}.

\bibitem[{\citenamefont{Bron et~al.}(1979)\citenamefont{Bron, Hesselink, {Van
  Poelgeest}, Zalmstra, Uitzinger, Verheul, Heyde, Waroquier, Vincx, and {Van
  Isacker}}}]{Bron1979NPA}
\bibinfo{author}{\bibfnamefont{J.}~\bibnamefont{Bron}},
  \bibinfo{author}{\bibfnamefont{W.}~\bibnamefont{Hesselink}},
  \bibinfo{author}{\bibfnamefont{A.}~\bibnamefont{{Van Poelgeest}}},
  \bibinfo{author}{\bibfnamefont{J.}~\bibnamefont{Zalmstra}},
  \bibinfo{author}{\bibfnamefont{M.}~\bibnamefont{Uitzinger}},
  \bibinfo{author}{\bibfnamefont{H.}~\bibnamefont{Verheul}},
  \bibinfo{author}{\bibfnamefont{K.}~\bibnamefont{Heyde}},
  \bibinfo{author}{\bibfnamefont{M.}~\bibnamefont{Waroquier}},
  \bibinfo{author}{\bibfnamefont{H.}~\bibnamefont{Vincx}}, \bibnamefont{and}
  \bibinfo{author}{\bibfnamefont{P.}~\bibnamefont{{Van Isacker}}},
  \bibinfo{journal}{Nuclear Physics A} \textbf{\bibinfo{volume}{318}},
  \bibinfo{pages}{335} (\bibinfo{year}{1979}), ISSN \bibinfo{issn}{0375-9474}.

\bibitem[{\citenamefont{Cheifetz et~al.}(1970)\citenamefont{Cheifetz, Jared,
  Thompson, and Wilhelmy}}]{Cheifetz1970PRL}
\bibinfo{author}{\bibfnamefont{E.}~\bibnamefont{Cheifetz}},
  \bibinfo{author}{\bibfnamefont{R.~C.} \bibnamefont{Jared}},
  \bibinfo{author}{\bibfnamefont{S.~G.} \bibnamefont{Thompson}},
  \bibnamefont{and} \bibinfo{author}{\bibfnamefont{J.~B.}
  \bibnamefont{Wilhelmy}}, \bibinfo{journal}{Phys. Rev. Lett.}
  \textbf{\bibinfo{volume}{25}}, \bibinfo{pages}{38} (\bibinfo{year}{1970}).

\bibitem[{\citenamefont{Federman and Pittel}(1977)}]{Federman1977PLB}
\bibinfo{author}{\bibfnamefont{P.}~\bibnamefont{Federman}} \bibnamefont{and}
  \bibinfo{author}{\bibfnamefont{S.}~\bibnamefont{Pittel}},
  \bibinfo{journal}{Physics Letters B} \textbf{\bibinfo{volume}{69}},
  \bibinfo{pages}{385} (\bibinfo{year}{1977}), ISSN \bibinfo{issn}{0370-2693}.

\bibitem[{\citenamefont{Federman and Pittel}(1979)}]{Federman1979prc}
\bibinfo{author}{\bibfnamefont{P.}~\bibnamefont{Federman}} \bibnamefont{and}
  \bibinfo{author}{\bibfnamefont{S.}~\bibnamefont{Pittel}},
  \bibinfo{journal}{Phys. Rev. C} \textbf{\bibinfo{volume}{20}},
  \bibinfo{pages}{820} (\bibinfo{year}{1979}).

\bibitem[{\citenamefont{Hager et~al.}(2007)\citenamefont{Hager, Jokinen,
  Elomaa, Eronen, Hakala, Kankainen, Rahaman, Rissanen, Moore, Rinta-Antila
  et~al.}}]{Hager2007NPA}
\bibinfo{author}{\bibfnamefont{U.}~\bibnamefont{Hager}},
  \bibinfo{author}{\bibfnamefont{A.}~\bibnamefont{Jokinen}},
  \bibinfo{author}{\bibfnamefont{V.-V.} \bibnamefont{Elomaa}},
  \bibinfo{author}{\bibfnamefont{T.}~\bibnamefont{Eronen}},
  \bibinfo{author}{\bibfnamefont{J.}~\bibnamefont{Hakala}},
  \bibinfo{author}{\bibfnamefont{A.}~\bibnamefont{Kankainen}},
  \bibinfo{author}{\bibfnamefont{S.}~\bibnamefont{Rahaman}},
  \bibinfo{author}{\bibfnamefont{J.}~\bibnamefont{Rissanen}},
  \bibinfo{author}{\bibfnamefont{I.}~\bibnamefont{Moore}},
  \bibinfo{author}{\bibfnamefont{S.}~\bibnamefont{Rinta-Antila}},
  \bibnamefont{et~al.}, \bibinfo{journal}{Nuclear Physics A}
  \textbf{\bibinfo{volume}{793}}, \bibinfo{pages}{20} (\bibinfo{year}{2007}),
  ISSN \bibinfo{issn}{0375-9474}.

\bibitem[{\citenamefont{Hamilton et~al.}(1974)\citenamefont{Hamilton, Ramayya,
  Pinkston, Ronningen, Garcia-Bermudez, Carter, Robinson, Kim, and
  Sayer}}]{Hamilton1974PRL}
\bibinfo{author}{\bibfnamefont{J.~H.} \bibnamefont{Hamilton}},
  \bibinfo{author}{\bibfnamefont{A.~V.} \bibnamefont{Ramayya}},
  \bibinfo{author}{\bibfnamefont{W.~T.} \bibnamefont{Pinkston}},
  \bibinfo{author}{\bibfnamefont{R.~M.} \bibnamefont{Ronningen}},
  \bibinfo{author}{\bibfnamefont{G.}~\bibnamefont{Garcia-Bermudez}},
  \bibinfo{author}{\bibfnamefont{H.~K.} \bibnamefont{Carter}},
  \bibinfo{author}{\bibfnamefont{R.~L.} \bibnamefont{Robinson}},
  \bibinfo{author}{\bibfnamefont{H.~J.} \bibnamefont{Kim}}, \bibnamefont{and}
  \bibinfo{author}{\bibfnamefont{R.~O.} \bibnamefont{Sayer}},
  \bibinfo{journal}{Phys. Rev. Lett.} \textbf{\bibinfo{volume}{32}},
  \bibinfo{pages}{239} (\bibinfo{year}{1974}).

\bibitem[{\citenamefont{Caurier et~al.}(2005)\citenamefont{Caurier,
  Mart\'{\i}nez-Pinedo, Nowacki, Poves, and Zuker}}]{Caurier2005RMP}
\bibinfo{author}{\bibfnamefont{E.}~\bibnamefont{Caurier}},
  \bibinfo{author}{\bibfnamefont{G.}~\bibnamefont{Mart\'{\i}nez-Pinedo}},
  \bibinfo{author}{\bibfnamefont{F.}~\bibnamefont{Nowacki}},
  \bibinfo{author}{\bibfnamefont{A.}~\bibnamefont{Poves}}, \bibnamefont{and}
  \bibinfo{author}{\bibfnamefont{A.~P.} \bibnamefont{Zuker}},
  \bibinfo{journal}{Rev. Mod. Phys.} \textbf{\bibinfo{volume}{77}},
  \bibinfo{pages}{427} (\bibinfo{year}{2005}).

\bibitem[{\citenamefont{Otsuka et~al.}(2001)\citenamefont{Otsuka, Honma,
  Mizusaki, Shimizu, and Utsuno}}]{Otsuka2001PPNP}
\bibinfo{author}{\bibfnamefont{T.}~\bibnamefont{Otsuka}},
  \bibinfo{author}{\bibfnamefont{M.}~\bibnamefont{Honma}},
  \bibinfo{author}{\bibfnamefont{T.}~\bibnamefont{Mizusaki}},
  \bibinfo{author}{\bibfnamefont{N.}~\bibnamefont{Shimizu}}, \bibnamefont{and}
  \bibinfo{author}{\bibfnamefont{Y.}~\bibnamefont{Utsuno}},
  \bibinfo{journal}{Progress in Particle and Nuclear Physics}
  \textbf{\bibinfo{volume}{47}}, \bibinfo{pages}{319} (\bibinfo{year}{2001}),
  ISSN \bibinfo{issn}{0146-6410}.

\bibitem[{\citenamefont{Nomura et~al.}(2016)\citenamefont{Nomura, Otsuka, and
  Isacker}}]{Nomura2016JPG}
\bibinfo{author}{\bibfnamefont{K.}~\bibnamefont{Nomura}},
  \bibinfo{author}{\bibfnamefont{T.}~\bibnamefont{Otsuka}}, \bibnamefont{and}
  \bibinfo{author}{\bibfnamefont{P.~V.} \bibnamefont{Isacker}},
  \bibinfo{journal}{Journal of Physics G: Nuclear and Particle Physics}
  \textbf{\bibinfo{volume}{43}}, \bibinfo{pages}{024008}
  (\bibinfo{year}{2016}).

\bibitem[{\citenamefont{Bender et~al.}(2003)\citenamefont{Bender, Heenen, and
  Reinhard}}]{Bender2003RMP}
\bibinfo{author}{\bibfnamefont{M.}~\bibnamefont{Bender}},
  \bibinfo{author}{\bibfnamefont{P.-H.} \bibnamefont{Heenen}},
  \bibnamefont{and} \bibinfo{author}{\bibfnamefont{P.-G.}
  \bibnamefont{Reinhard}}, \bibinfo{journal}{Rev. Mod. Phys.}
  \textbf{\bibinfo{volume}{75}}, \bibinfo{pages}{121} (\bibinfo{year}{2003}).

\bibitem[{\citenamefont{Robledo et~al.}(2018)\citenamefont{Robledo, Rodríguez,
  and Rodríguez-Guzmán}}]{Robledo2019JPG}
\bibinfo{author}{\bibfnamefont{L.~M.} \bibnamefont{Robledo}},
  \bibinfo{author}{\bibfnamefont{T.~R.} \bibnamefont{Rodríguez}},
  \bibnamefont{and} \bibinfo{author}{\bibfnamefont{R.~R.}
  \bibnamefont{Rodríguez-Guzmán}}, \bibinfo{journal}{Journal of Physics G:
  Nuclear and Particle Physics} \textbf{\bibinfo{volume}{46}},
  \bibinfo{pages}{013001} (\bibinfo{year}{2018}).

\bibitem[{\citenamefont{Nikšić et~al.}(2011)\citenamefont{Nikšić, Vretenar,
  and Ring}}]{NIKSIC2011PPNP}
\bibinfo{author}{\bibfnamefont{T.}~\bibnamefont{Nikšić}},
  \bibinfo{author}{\bibfnamefont{D.}~\bibnamefont{Vretenar}}, \bibnamefont{and}
  \bibinfo{author}{\bibfnamefont{P.}~\bibnamefont{Ring}},
  \bibinfo{journal}{Progress in Particle and Nuclear Physics}
  \textbf{\bibinfo{volume}{66}}, \bibinfo{pages}{519} (\bibinfo{year}{2011}),
  ISSN \bibinfo{issn}{0146-6410}.

\bibitem[{\citenamefont{Sagawa et~al.}(2004)\citenamefont{Sagawa, Zhou, Zhang,
  and Suzuki}}]{Sagawa04}
\bibinfo{author}{\bibfnamefont{H.}~\bibnamefont{Sagawa}},
  \bibinfo{author}{\bibfnamefont{X.~R.} \bibnamefont{Zhou}},
  \bibinfo{author}{\bibfnamefont{X.~Z.} \bibnamefont{Zhang}}, \bibnamefont{and}
  \bibinfo{author}{\bibfnamefont{T.}~\bibnamefont{Suzuki}},
  \bibinfo{journal}{Phys. Rev. C} \textbf{\bibinfo{volume}{70}},
  \bibinfo{pages}{054316} (\bibinfo{year}{2004}).

\bibitem[{\citenamefont{Li et~al.}(2013)\citenamefont{Li, Hiyama, Zhou, and
  Sagawa}}]{Lia2013prc}
\bibinfo{author}{\bibfnamefont{A.}~\bibnamefont{Li}},
  \bibinfo{author}{\bibfnamefont{E.}~\bibnamefont{Hiyama}},
  \bibinfo{author}{\bibfnamefont{X.-R.} \bibnamefont{Zhou}}, \bibnamefont{and}
  \bibinfo{author}{\bibfnamefont{H.}~\bibnamefont{Sagawa}},
  \bibinfo{journal}{Phys. Rev. C} \textbf{\bibinfo{volume}{87}},
  \bibinfo{pages}{014333} (\bibinfo{year}{2013}).

\bibitem[{\citenamefont{Rodr{\'\i}guez-Guzm{\'a}n
  et~al.}(2003)\citenamefont{Rodr{\'\i}guez-Guzm{\'a}n, Egido, and
  Robledo}}]{Rodriguez2003}
\bibinfo{author}{\bibfnamefont{R.}~\bibnamefont{Rodr{\'\i}guez-Guzm{\'a}n}},
  \bibinfo{author}{\bibfnamefont{J.}~\bibnamefont{Egido}}, \bibnamefont{and}
  \bibinfo{author}{\bibfnamefont{L.}~\bibnamefont{Robledo}},
  \bibinfo{journal}{The European Physical Journal A-Hadrons and Nuclei}
  \textbf{\bibinfo{volume}{17}}, \bibinfo{pages}{37} (\bibinfo{year}{2003}).

\bibitem[{\citenamefont{Feliciello and Nagae}(2015)}]{Feliciello2015}
\bibinfo{author}{\bibfnamefont{A.}~\bibnamefont{Feliciello}} \bibnamefont{and}
  \bibinfo{author}{\bibfnamefont{T.}~\bibnamefont{Nagae}},
  \bibinfo{journal}{Reports on Progress in Physics}
  \textbf{\bibinfo{volume}{78}}, \bibinfo{pages}{096301}
  (\bibinfo{year}{2015}).

\bibitem[{\citenamefont{Gal et~al.}(2016)\citenamefont{Gal, Hungerford, and
  Millener}}]{Gal16}
\bibinfo{author}{\bibfnamefont{A.}~\bibnamefont{Gal}},
  \bibinfo{author}{\bibfnamefont{E.~V.} \bibnamefont{Hungerford}},
  \bibnamefont{and} \bibinfo{author}{\bibfnamefont{D.~J.}
  \bibnamefont{Millener}}, \bibinfo{journal}{Rev. Mod. Phys.}
  \textbf{\bibinfo{volume}{88}}, \bibinfo{pages}{035004}
  (\bibinfo{year}{2016}).

\bibitem[{\citenamefont{Schulze and Hiyama}(2014)}]{Schulze2014PRC}
\bibinfo{author}{\bibfnamefont{H.-J.} \bibnamefont{Schulze}} \bibnamefont{and}
  \bibinfo{author}{\bibfnamefont{E.}~\bibnamefont{Hiyama}},
  \bibinfo{journal}{Phys. Rev. C} \textbf{\bibinfo{volume}{90}},
  \bibinfo{pages}{047301} (\bibinfo{year}{2014}).

\bibitem[{\citenamefont{{Xue} et~al.}(2022)\citenamefont{{Xue}, {Chen}, {Zhou},
  {Cheng}, and {Schulze}}}]{Xue22}
\bibinfo{author}{\bibfnamefont{H.-T.} \bibnamefont{{Xue}}},
  \bibinfo{author}{\bibfnamefont{Q.~B.} \bibnamefont{{Chen}}},
  \bibinfo{author}{\bibfnamefont{X.-R.} \bibnamefont{{Zhou}}},
  \bibinfo{author}{\bibfnamefont{Y.~Y.} \bibnamefont{{Cheng}}},
  \bibnamefont{and} \bibinfo{author}{\bibfnamefont{H.~J.}
  \bibnamefont{{Schulze}}}, \bibinfo{journal}{\prc}
  \textbf{\bibinfo{volume}{106}}, \bibinfo{eid}{044306} (\bibinfo{year}{2022}).

\bibitem[{\citenamefont{Xue et~al.}(2023)\citenamefont{Xue, Chen, Chen, Luo,
  Schulze, and Zhou}}]{Xue23}
\bibinfo{author}{\bibfnamefont{H.-T.} \bibnamefont{Xue}},
  \bibinfo{author}{\bibfnamefont{Y.-F.} \bibnamefont{Chen}},
  \bibinfo{author}{\bibfnamefont{Q.~B.} \bibnamefont{Chen}},
  \bibinfo{author}{\bibfnamefont{Y.~A.} \bibnamefont{Luo}},
  \bibinfo{author}{\bibfnamefont{H.-J.} \bibnamefont{Schulze}},
  \bibnamefont{and} \bibinfo{author}{\bibfnamefont{X.-R.} \bibnamefont{Zhou}},
  \bibinfo{journal}{\prc} \textbf{\bibinfo{volume}{107}},
  \bibinfo{pages}{044317} (\bibinfo{year}{2023}).

\bibitem[{\citenamefont{Chen et~al.}(2022)\citenamefont{Chen, Zhou, Chen, and
  Cheng}}]{chen_2022_epja}
\bibinfo{author}{\bibfnamefont{Y.-F.} \bibnamefont{Chen}},
  \bibinfo{author}{\bibfnamefont{X.-R.} \bibnamefont{Zhou}},
  \bibinfo{author}{\bibfnamefont{Q.}~\bibnamefont{Chen}}, \bibnamefont{and}
  \bibinfo{author}{\bibfnamefont{Y.-Y.} \bibnamefont{Cheng}},
  \bibinfo{journal}{The European Physical Journal A}
  \textbf{\bibinfo{volume}{58}}, \bibinfo{pages}{1} (\bibinfo{year}{2022}).

\bibitem[{\citenamefont{Guo et~al.}(2022)\citenamefont{Guo, Chen, Zhou, Chen,
  and Schulze}}]{Guo2022prc}
\bibinfo{author}{\bibfnamefont{J.}~\bibnamefont{Guo}},
  \bibinfo{author}{\bibfnamefont{C.~F.} \bibnamefont{Chen}},
  \bibinfo{author}{\bibfnamefont{X.-R.} \bibnamefont{Zhou}},
  \bibinfo{author}{\bibfnamefont{Q.~B.} \bibnamefont{Chen}}, \bibnamefont{and}
  \bibinfo{author}{\bibfnamefont{H.-J.} \bibnamefont{Schulze}},
  \bibinfo{journal}{Phys. Rev. C} \textbf{\bibinfo{volume}{105}},
  \bibinfo{pages}{034322} (\bibinfo{year}{2022}).

\bibitem[{\citenamefont{Liu et~al.}(2023)\citenamefont{Liu, Chen, Chen, Xue,
  Schulze, and Zhou}}]{Liu2024prc}
\bibinfo{author}{\bibfnamefont{Y.-X.} \bibnamefont{Liu}},
  \bibinfo{author}{\bibfnamefont{C.~F.} \bibnamefont{Chen}},
  \bibinfo{author}{\bibfnamefont{Q.~B.} \bibnamefont{Chen}},
  \bibinfo{author}{\bibfnamefont{H.-T.} \bibnamefont{Xue}},
  \bibinfo{author}{\bibfnamefont{H.-J.} \bibnamefont{Schulze}},
  \bibnamefont{and} \bibinfo{author}{\bibfnamefont{X.-R.} \bibnamefont{Zhou}},
  \bibinfo{journal}{Phys. Rev. C} \textbf{\bibinfo{volume}{108}},
  \bibinfo{pages}{064312} (\bibinfo{year}{2023}).

\bibitem[{\citenamefont{Li et~al.}(2024)\citenamefont{Li, Chen, Zhou, and
  Ren}}]{Li2024prc}
\bibinfo{author}{\bibfnamefont{X.}~\bibnamefont{Li}},
  \bibinfo{author}{\bibfnamefont{C.~F.} \bibnamefont{Chen}},
  \bibinfo{author}{\bibfnamefont{X.-R.} \bibnamefont{Zhou}}, \bibnamefont{and}
  \bibinfo{author}{\bibfnamefont{Z.}~\bibnamefont{Ren}},
  \bibinfo{journal}{Phys. Rev. C} \textbf{\bibinfo{volume}{109}},
  \bibinfo{pages}{064301} (\bibinfo{year}{2024}).

\bibitem[{\citenamefont{Cui et~al.}(2017)\citenamefont{Cui, Zhou, Guo, and
  Schulze}}]{Cui2017prc}
\bibinfo{author}{\bibfnamefont{J.-W.} \bibnamefont{Cui}},
  \bibinfo{author}{\bibfnamefont{X.-R.} \bibnamefont{Zhou}},
  \bibinfo{author}{\bibfnamefont{L.-X.} \bibnamefont{Guo}}, \bibnamefont{and}
  \bibinfo{author}{\bibfnamefont{H.-J.} \bibnamefont{Schulze}},
  \bibinfo{journal}{Phys. Rev. C} \textbf{\bibinfo{volume}{95}},
  \bibinfo{pages}{024323} (\bibinfo{year}{2017}).

\bibitem[{\citenamefont{Cui et~al.}(2022)\citenamefont{Cui, Wang, and
  Zhou}}]{Cui2022cpc}
\bibinfo{author}{\bibfnamefont{J.-W.} \bibnamefont{Cui}},
  \bibinfo{author}{\bibfnamefont{R.}~\bibnamefont{Wang}}, \bibnamefont{and}
  \bibinfo{author}{\bibfnamefont{X.-R.} \bibnamefont{Zhou}},
  \bibinfo{journal}{Chinese Physics C} \textbf{\bibinfo{volume}{46}},
  \bibinfo{pages}{074109} (\bibinfo{year}{2022}).

\bibitem[{\citenamefont{Mei et~al.}(2015)\citenamefont{Mei, Hagino, Yao, and
  Motoba}}]{Mei2015prc}
\bibinfo{author}{\bibfnamefont{H.}~\bibnamefont{Mei}},
  \bibinfo{author}{\bibfnamefont{K.}~\bibnamefont{Hagino}},
  \bibinfo{author}{\bibfnamefont{J.~M.} \bibnamefont{Yao}}, \bibnamefont{and}
  \bibinfo{author}{\bibfnamefont{T.}~\bibnamefont{Motoba}},
  \bibinfo{journal}{Phys. Rev. C} \textbf{\bibinfo{volume}{91}},
  \bibinfo{pages}{064305} (\bibinfo{year}{2015}).

\bibitem[{\citenamefont{Xia et~al.}(2019)\citenamefont{Xia, Wu, Mei, and
  Yao}}]{XJH2019}
\bibinfo{author}{\bibfnamefont{H.}~\bibnamefont{Xia}},
  \bibinfo{author}{\bibfnamefont{X.}~\bibnamefont{Wu}},
  \bibinfo{author}{\bibfnamefont{H.}~\bibnamefont{Mei}}, \bibnamefont{and}
  \bibinfo{author}{\bibfnamefont{J.}~\bibnamefont{Yao}}, \bibinfo{journal}{Sci.
  China-Phys. Mech. Astron.} \textbf{\bibinfo{volume}{62}},
  \bibinfo{pages}{42011} (\bibinfo{year}{2019}).

\bibitem[{\citenamefont{Xue et~al.}(2024)\citenamefont{Xue, Chen, Cui, Chen,
  Schulze, and Zhou}}]{Xue2024prc}
\bibinfo{author}{\bibfnamefont{H.-T.} \bibnamefont{Xue}},
  \bibinfo{author}{\bibfnamefont{Q.~B.} \bibnamefont{Chen}},
  \bibinfo{author}{\bibfnamefont{J.-W.} \bibnamefont{Cui}},
  \bibinfo{author}{\bibfnamefont{C.-F.} \bibnamefont{Chen}},
  \bibinfo{author}{\bibfnamefont{H.-J.} \bibnamefont{Schulze}},
  \bibnamefont{and} \bibinfo{author}{\bibfnamefont{X.-R.} \bibnamefont{Zhou}},
  \bibinfo{journal}{Phys. Rev. C} \textbf{\bibinfo{volume}{109}},
  \bibinfo{pages}{024324} (\bibinfo{year}{2024}).

\bibitem[{\citenamefont{Motoba et~al.}(1983)\citenamefont{Motoba, Bandō, and
  Ikeda}}]{Motoba1983PTP}
\bibinfo{author}{\bibfnamefont{T.}~\bibnamefont{Motoba}},
  \bibinfo{author}{\bibfnamefont{H.}~\bibnamefont{Bandō}}, \bibnamefont{and}
  \bibinfo{author}{\bibfnamefont{K.}~\bibnamefont{Ikeda}},
  \bibinfo{journal}{Progress of Theoretical Physics}
  \textbf{\bibinfo{volume}{70}}, \bibinfo{pages}{189} (\bibinfo{year}{1983}),
  ISSN \bibinfo{issn}{0033-068X}.

\bibitem[{\citenamefont{Hiyama et~al.}(1999)\citenamefont{Hiyama, Kamimura,
  Miyazaki, and Motoba}}]{Hiyama1999}
\bibinfo{author}{\bibfnamefont{E.}~\bibnamefont{Hiyama}},
  \bibinfo{author}{\bibfnamefont{M.}~\bibnamefont{Kamimura}},
  \bibinfo{author}{\bibfnamefont{K.}~\bibnamefont{Miyazaki}}, \bibnamefont{and}
  \bibinfo{author}{\bibfnamefont{T.}~\bibnamefont{Motoba}},
  \bibinfo{journal}{Phys. Rev. C} \textbf{\bibinfo{volume}{59}},
  \bibinfo{pages}{2351} (\bibinfo{year}{1999}).

\bibitem[{\citenamefont{Hiyama et~al.}(1997)\citenamefont{Hiyama, Kamimura,
  Motoba, Yamada, and Yamamoto}}]{Hiyama1997PTP}
\bibinfo{author}{\bibfnamefont{E.}~\bibnamefont{Hiyama}},
  \bibinfo{author}{\bibfnamefont{M.}~\bibnamefont{Kamimura}},
  \bibinfo{author}{\bibfnamefont{T.}~\bibnamefont{Motoba}},
  \bibinfo{author}{\bibfnamefont{T.}~\bibnamefont{Yamada}}, \bibnamefont{and}
  \bibinfo{author}{\bibfnamefont{Y.}~\bibnamefont{Yamamoto}},
  \bibinfo{journal}{Progress of Theoretical Physics}
  \textbf{\bibinfo{volume}{97}}, \bibinfo{pages}{881} (\bibinfo{year}{1997}),
  ISSN \bibinfo{issn}{0033-068X},
  \eprint{https://academic.oup.com/ptp/article-pdf/97/6/881/5339280/97-6-881.pdf}.

\bibitem[{\citenamefont{Yao et~al.}(2011)\citenamefont{Yao, Li, Hagino, Win,
  Zhang, and Meng}}]{Yao2011NPA}
\bibinfo{author}{\bibfnamefont{J.}~\bibnamefont{Yao}},
  \bibinfo{author}{\bibfnamefont{Z.}~\bibnamefont{Li}},
  \bibinfo{author}{\bibfnamefont{K.}~\bibnamefont{Hagino}},
  \bibinfo{author}{\bibfnamefont{M.}~\bibnamefont{Win}},
  \bibinfo{author}{\bibfnamefont{Y.}~\bibnamefont{Zhang}}, \bibnamefont{and}
  \bibinfo{author}{\bibfnamefont{J.}~\bibnamefont{Meng}},
  \bibinfo{journal}{Nuclear Physics A} \textbf{\bibinfo{volume}{868-869}},
  \bibinfo{pages}{12} (\bibinfo{year}{2011}), ISSN \bibinfo{issn}{0375-9474}.

\bibitem[{\citenamefont{Hagino et~al.}(2013)\citenamefont{Hagino, Yao, Minato,
  Li, and {Thi Win}}}]{Hagino2013NPA}
\bibinfo{author}{\bibfnamefont{K.}~\bibnamefont{Hagino}},
  \bibinfo{author}{\bibfnamefont{J.}~\bibnamefont{Yao}},
  \bibinfo{author}{\bibfnamefont{F.}~\bibnamefont{Minato}},
  \bibinfo{author}{\bibfnamefont{Z.}~\bibnamefont{Li}}, \bibnamefont{and}
  \bibinfo{author}{\bibfnamefont{M.}~\bibnamefont{{Thi Win}}},
  \bibinfo{journal}{Nuclear Physics A} \textbf{\bibinfo{volume}{914}},
  \bibinfo{pages}{151} (\bibinfo{year}{2013}), ISSN \bibinfo{issn}{0375-9474},
  \bibinfo{note}{xI International Conference on Hypernuclear and Strange
  Particle Physics (HYP2012)}.

\bibitem[{\citenamefont{Vretenar et~al.}(1998)\citenamefont{Vretenar, P\"oschl,
  Lalazissis, and Ring}}]{Vretenar1998prc}
\bibinfo{author}{\bibfnamefont{D.}~\bibnamefont{Vretenar}},
  \bibinfo{author}{\bibfnamefont{W.}~\bibnamefont{P\"oschl}},
  \bibinfo{author}{\bibfnamefont{G.~A.} \bibnamefont{Lalazissis}},
  \bibnamefont{and} \bibinfo{author}{\bibfnamefont{P.}~\bibnamefont{Ring}},
  \bibinfo{journal}{Phys. Rev. C} \textbf{\bibinfo{volume}{57}},
  \bibinfo{pages}{R1060} (\bibinfo{year}{1998}).

\bibitem[{\citenamefont{Zhou et~al.}(2008)\citenamefont{Zhou, Polls, Schulze,
  and Vida\~na}}]{Zhou2008prc}
\bibinfo{author}{\bibfnamefont{X.-R.} \bibnamefont{Zhou}},
  \bibinfo{author}{\bibfnamefont{A.}~\bibnamefont{Polls}},
  \bibinfo{author}{\bibfnamefont{H.-J.} \bibnamefont{Schulze}},
  \bibnamefont{and} \bibinfo{author}{\bibfnamefont{I.}~\bibnamefont{Vida\~na}},
  \bibinfo{journal}{Phys. Rev. C} \textbf{\bibinfo{volume}{78}},
  \bibinfo{pages}{054306} (\bibinfo{year}{2008}).

\bibitem[{\citenamefont{L{\"u} et~al.}(2003)\citenamefont{L{\"u}, Meng, Zhang,
  and Zhou}}]{Lu2003epja}
\bibinfo{author}{\bibfnamefont{H.}~\bibnamefont{L{\"u}}},
  \bibinfo{author}{\bibfnamefont{J.}~\bibnamefont{Meng}},
  \bibinfo{author}{\bibfnamefont{S.}~\bibnamefont{Zhang}}, \bibnamefont{and}
  \bibinfo{author}{\bibfnamefont{S.-G.} \bibnamefont{Zhou}},
  \bibinfo{journal}{The European Physical Journal A-Hadrons and Nuclei}
  \textbf{\bibinfo{volume}{17}}, \bibinfo{pages}{19} (\bibinfo{year}{2003}).

\bibitem[{\citenamefont{Hiyama et~al.}(1996)\citenamefont{Hiyama, Kamimura,
  Motoba, Yamada, and Yamamoto}}]{Hiyama1996prc}
\bibinfo{author}{\bibfnamefont{E.}~\bibnamefont{Hiyama}},
  \bibinfo{author}{\bibfnamefont{M.}~\bibnamefont{Kamimura}},
  \bibinfo{author}{\bibfnamefont{T.}~\bibnamefont{Motoba}},
  \bibinfo{author}{\bibfnamefont{T.}~\bibnamefont{Yamada}}, \bibnamefont{and}
  \bibinfo{author}{\bibfnamefont{Y.}~\bibnamefont{Yamamoto}},
  \bibinfo{journal}{Phys. Rev. C} \textbf{\bibinfo{volume}{53}},
  \bibinfo{pages}{2075} (\bibinfo{year}{1996}).

\bibitem[{\citenamefont{Xia et~al.}(2023)\citenamefont{Xia, Wu, Mei, and
  Yao}}]{XJH2023}
\bibinfo{author}{\bibfnamefont{H.}~\bibnamefont{Xia}},
  \bibinfo{author}{\bibfnamefont{X.}~\bibnamefont{Wu}},
  \bibinfo{author}{\bibfnamefont{H.}~\bibnamefont{Mei}}, \bibnamefont{and}
  \bibinfo{author}{\bibfnamefont{J.}~\bibnamefont{Yao}}, \bibinfo{journal}{Sci.
  China-Phys. Mech. Astron.} \textbf{\bibinfo{volume}{66}},
  \bibinfo{pages}{252011} (\bibinfo{year}{2023}).

\bibitem[{\citenamefont{Chen et~al.}(2021)\citenamefont{Chen, Sun, Li, and
  Sun}}]{Chen2021SCP}
\bibinfo{author}{\bibfnamefont{C.}~\bibnamefont{Chen}},
  \bibinfo{author}{\bibfnamefont{Q.-K.} \bibnamefont{Sun}},
  \bibinfo{author}{\bibfnamefont{Y.-X.} \bibnamefont{Li}}, \bibnamefont{and}
  \bibinfo{author}{\bibfnamefont{T.-T.} \bibnamefont{Sun}},
  \bibinfo{journal}{Science China Physics, Mechanics \& Astronomy}
  \textbf{\bibinfo{volume}{64}}, \bibinfo{pages}{282011}
  (\bibinfo{year}{2021}).

\bibitem[{\citenamefont{Bender et~al.}(2000)\citenamefont{Bender, Rutz,
  Reinhard, and Maruhn}}]{Bender2000epja}
\bibinfo{author}{\bibfnamefont{M.}~\bibnamefont{Bender}},
  \bibinfo{author}{\bibfnamefont{K.}~\bibnamefont{Rutz}},
  \bibinfo{author}{\bibfnamefont{P.~G.} \bibnamefont{Reinhard}},
  \bibnamefont{and} \bibinfo{author}{\bibfnamefont{J.~A.}
  \bibnamefont{Maruhn}}, \bibinfo{journal}{The European Physical Journal A}
  \textbf{\bibinfo{volume}{8}}, \bibinfo{pages}{59} (\bibinfo{year}{2000}).

\bibitem[{\citenamefont{Ring and Schuck}(1980)}]{Peter1980}
\bibinfo{author}{\bibfnamefont{P.}~\bibnamefont{Ring}} \bibnamefont{and}
  \bibinfo{author}{\bibfnamefont{P.}~\bibnamefont{Schuck}},
  \emph{\bibinfo{title}{The nuclear many body problem}}
  (\bibinfo{publisher}{Springer Verlag, Berlin}, \bibinfo{year}{1980}).

\bibitem[{\citenamefont{Bonche et~al.}(1990)\citenamefont{Bonche, Dobaczewski,
  Flocard, Heenen, and Meyer}}]{Bonche1990npa}
\bibinfo{author}{\bibfnamefont{P.}~\bibnamefont{Bonche}},
  \bibinfo{author}{\bibfnamefont{J.}~\bibnamefont{Dobaczewski}},
  \bibinfo{author}{\bibfnamefont{H.}~\bibnamefont{Flocard}},
  \bibinfo{author}{\bibfnamefont{P.-H.} \bibnamefont{Heenen}},
  \bibnamefont{and} \bibinfo{author}{\bibfnamefont{J.}~\bibnamefont{Meyer}},
  \bibinfo{journal}{Nuclear Physics A} \textbf{\bibinfo{volume}{510}},
  \bibinfo{pages}{466} (\bibinfo{year}{1990}), ISSN \bibinfo{issn}{0375-9474}.

\bibitem[{\citenamefont{Bender et~al.}(2006)\citenamefont{Bender, Bonche, and
  Heenen}}]{Bender2006prc}
\bibinfo{author}{\bibfnamefont{M.}~\bibnamefont{Bender}},
  \bibinfo{author}{\bibfnamefont{P.}~\bibnamefont{Bonche}}, \bibnamefont{and}
  \bibinfo{author}{\bibfnamefont{P.-H.} \bibnamefont{Heenen}},
  \bibinfo{journal}{Phys. Rev. C} \textbf{\bibinfo{volume}{74}},
  \bibinfo{pages}{024312} (\bibinfo{year}{2006}).

\bibitem[{\citenamefont{{Dobaczewski} et~al.}(2009)}]{Dobaczewski09}
\bibinfo{author}{\bibfnamefont{J.}~\bibnamefont{{Dobaczewski}}}
  \bibnamefont{et~al.}, \bibinfo{journal}{Computer Physics Communications}
  \textbf{\bibinfo{volume}{180}}, \bibinfo{pages}{2361} (\bibinfo{year}{2009}),
  \eprint{0903.1020}.

\bibitem[{\citenamefont{Sumi et~al.}(2012)\citenamefont{Sumi, Minomo, Tagami,
  Kimura, Matsumoto, Ogata, Shimizu, and Yahiro}}]{Sumi2012prc}
\bibinfo{author}{\bibfnamefont{T.}~\bibnamefont{Sumi}},
  \bibinfo{author}{\bibfnamefont{K.}~\bibnamefont{Minomo}},
  \bibinfo{author}{\bibfnamefont{S.}~\bibnamefont{Tagami}},
  \bibinfo{author}{\bibfnamefont{M.}~\bibnamefont{Kimura}},
  \bibinfo{author}{\bibfnamefont{T.}~\bibnamefont{Matsumoto}},
  \bibinfo{author}{\bibfnamefont{K.}~\bibnamefont{Ogata}},
  \bibinfo{author}{\bibfnamefont{Y.~R.} \bibnamefont{Shimizu}},
  \bibnamefont{and} \bibinfo{author}{\bibfnamefont{M.}~\bibnamefont{Yahiro}},
  \bibinfo{journal}{Phys. Rev. C} \textbf{\bibinfo{volume}{85}},
  \bibinfo{pages}{064613} (\bibinfo{year}{2012}).

\bibitem[{NuD()}]{NuDat}
\emph{\bibinfo{title}{National nuclear data center}},
  \bibinfo{note}{\url{https://www.nndc.bnl.gov/nudat3/}}.

\bibitem[{\citenamefont{Hilaire and Girod}(2007)}]{hilaire2007epja}
\bibinfo{author}{\bibfnamefont{S.}~\bibnamefont{Hilaire}} \bibnamefont{and}
  \bibinfo{author}{\bibfnamefont{M.}~\bibnamefont{Girod}},
  \bibinfo{journal}{The European Physical Journal A}
  \textbf{\bibinfo{volume}{33}}, \bibinfo{pages}{237} (\bibinfo{year}{2007}).

\bibitem[{phy()}]{phynu}
\emph{\bibinfo{title}{{HFB} results based on the gogny force}},
  \bibinfo{note}{\url{https://www-phynu.cea.fr/science_en_ligne/carte_potentiels_microscopiques/carte_potentiel_nucleaire_eng.htm}}.

\bibitem[{\citenamefont{Obertelli et~al.}(2005)\citenamefont{Obertelli, P\'eru,
  Delaroche, Gillibert, Girod, and Goutte}}]{Obertelli2005prc}
\bibinfo{author}{\bibfnamefont{A.}~\bibnamefont{Obertelli}},
  \bibinfo{author}{\bibfnamefont{S.}~\bibnamefont{P\'eru}},
  \bibinfo{author}{\bibfnamefont{J.~P.} \bibnamefont{Delaroche}},
  \bibinfo{author}{\bibfnamefont{A.}~\bibnamefont{Gillibert}},
  \bibinfo{author}{\bibfnamefont{M.}~\bibnamefont{Girod}}, \bibnamefont{and}
  \bibinfo{author}{\bibfnamefont{H.}~\bibnamefont{Goutte}},
  \bibinfo{journal}{Phys. Rev. C} \textbf{\bibinfo{volume}{71}},
  \bibinfo{pages}{024304} (\bibinfo{year}{2005}).

\bibitem[{\citenamefont{Rodr{\'\i}guez-Guzm{\'a}n
  et~al.}(2002)\citenamefont{Rodr{\'\i}guez-Guzm{\'a}n, Egido, and
  Robledo}}]{Rodriguez2002NPA}
\bibinfo{author}{\bibfnamefont{R.}~\bibnamefont{Rodr{\'\i}guez-Guzm{\'a}n}},
  \bibinfo{author}{\bibfnamefont{J.}~\bibnamefont{Egido}}, \bibnamefont{and}
  \bibinfo{author}{\bibfnamefont{L.}~\bibnamefont{Robledo}},
  \bibinfo{journal}{Nuclear Physics A} \textbf{\bibinfo{volume}{709}},
  \bibinfo{pages}{201} (\bibinfo{year}{2002}).

\bibitem[{\citenamefont{Cui et~al.}(2015)\citenamefont{Cui, Zhou, and
  Schulze}}]{Cui2015prc}
\bibinfo{author}{\bibfnamefont{J.-W.} \bibnamefont{Cui}},
  \bibinfo{author}{\bibfnamefont{X.-R.} \bibnamefont{Zhou}}, \bibnamefont{and}
  \bibinfo{author}{\bibfnamefont{H.-J.} \bibnamefont{Schulze}},
  \bibinfo{journal}{Phys. Rev. C} \textbf{\bibinfo{volume}{91}},
  \bibinfo{pages}{054306} (\bibinfo{year}{2015}).

\end{thebibliography}

\end{CJK*}

\end{document}